\PassOptionsToPackage{unicode}{hyperref}
\PassOptionsToPackage{hyphens}{url}
\documentclass[11pt, a4paper]{article}
\usepackage{xcolor}
\usepackage{amsmath,amssymb}
\usepackage{needspace}
\usepackage{tikz}
\usepackage{algorithm}
\usepackage{algorithmic}
\usepackage{mdframed}
\usepackage{enumitem}
\usepackage[margin=1in]{geometry}
\usepackage{iftex}
\ifPDFTeX
  \usepackage[T1]{fontenc}
  \usepackage[utf8]{inputenc}
  \usepackage{textcomp} 
\else 
  \usepackage{unicode-math} 
  \defaultfontfeatures{Scale=MatchLowercase}
  \defaultfontfeatures[\rmfamily]{Ligatures=TeX,Scale=1}
\fi
\usepackage{lmodern}
\IfFileExists{upquote.sty}{\usepackage{upquote}}{}
\IfFileExists{microtype.sty}{
  \usepackage[]{microtype}
  \UseMicrotypeSet[protrusion]{basicmath} 
}{}
\makeatletter
\@ifundefined{KOMAClassName}{
  \IfFileExists{parskip.sty}{%
    \usepackage{parskip}
  }{
    \setlength{\parindent}{0pt}
    \setlength{\parskip}{6pt plus 2pt minus 1pt}}
}{
  \KOMAoptions{parskip=half}}
\makeatother
\usepackage{color}
\usepackage{fancyvrb}

\DefineVerbatimEnvironment{Highlighting}{Verbatim}{commandchars=\\\{\}}

\usepackage{longtable,booktabs,array}
\usepackage{calc} 
\usepackage{etoolbox}
\makeatletter
\patchcmd\longtable{\par}{\if@noskipsec\mbox{}\fi\par}{}{}
\makeatother
\IfFileExists{footnotehyper.sty}{\usepackage{footnotehyper}}{\usepackage{footnote}}
\makesavenoteenv{longtable}
\providecommand{\tightlist}{%
  \setlength{\itemsep}{0pt}\setlength{\parskip}{0pt}}
\usepackage{listings}
\usepackage{bookmark}
\IfFileExists{xurl.sty}{\usepackage{xurl}}{} 
\hypersetup{
  hidelinks,
  pdfcreator={LaTeX via pandoc}}

\title{zk-X509: Privacy-Preserving On-Chain Identity\\from Legacy PKI via Zero-Knowledge Proofs}

\author{Yeongju Bak\\
Tokamak Network\\
\texttt{zena@tokamak.network}}

\date{March 31, 2026\\{\small arXiv preprint --- v2}}

\begin{document}

\maketitle

\begin{abstract}

Public blockchains impose an inherent tension between regulatory
compliance and user privacy. Existing on-chain identity solutions
require either centralized KYC attestors---introducing single points of
failure and metadata leakage---specialized hardware such as NFC readers
or biometric scanners, or Decentralized Identifier (DID) frameworks that
necessitate building entirely new credential issuance infrastructure.
Meanwhile, over four billion active X.509 digital certificates (estimated from Certificate Transparency logs and TLS survey data)
constitute a globally deployed, government-grade trust infrastructure
that remains unexploited for decentralized identity.

This paper presents zk-X509, a software-based, privacy-preserving
identity system that bridges legacy Public Key Infrastructure (PKI) with
public ledgers via a RISC-V-based zero-knowledge virtual machine (zkVM).
The system enables users to prove ownership and validity of standard
X.509 certificates---issued by any national PKI or corporate
CA---without revealing private keys or personal identifiers. A key
architectural distinction is that the private key never enters the ZK
circuit; ownership is instead proven via signature verification
delegated to the OS keychain (macOS Security.framework, Windows CNG). The
circuit performs six verification checks: (1) full certificate chain validity to
a trusted root, (2) temporal validity, (3) signature-based key
ownership, (4) trustless CRL revocation checking with CA signature
verification inside the zkVM, (5) binding to a specific blockchain
address, and (6) configurable nullifier generation for Sybil resistance.
Thirteen public values are committed, including a nullifier, a CA Merkle
root that hides the issuing CA's identity, the certificate's expiry
timestamp, and four optional selective disclosure hashes. On-chain
verification status automatically lapses upon certificate expiry.

We formalize the security model under a Dolev-Yao adversary and
establish eight properties via game-based definitions: unforgeability,
unlinkability, cross-service unlinkability, cross-chain replay resistance, double-registration resistance, front-running immunity,
CA-membership hiding, and non-transferability, with explicit reductions to
sEUF-CMA signature security, SHA-256 modeled as a random oracle, and ZK
soundness. The SP1 zkVM implementation achieves 11.8 million cycles for
single-level ECDSA P-256 verification (17.4M for RSA-2048), with
on-chain verification costing approximately 300,000 gas under Groth16.
By leveraging certificates already deployed at scale across multiple
jurisdictions, zk-X509 enables adoption without new trust establishment,
providing a complementary approach to DID-based systems for integrating
existing trust anchors into decentralized applications.
\end{abstract}


\vspace{1em}\hrule\vspace{1em}

\section{Introduction}\label{introduction}

\subsection{Motivation}\label{motivation}

Digital identity verification on blockchain platforms faces a
fundamental tension between \textbf{transparency} and \textbf{privacy}.
Public blockchains provide immutable, auditable records, yet this same
transparency renders them unsuitable for storing personal identity data
such as names, national IDs, or certificate contents. Recent regulatory
actions---including OFAC sanctions on privacy-preserving protocols
\cite{ref1}---have intensified the demand for decentralized ``Proof of
Personhood'' (PoP) systems that satisfy compliance requirements without
sacrificing user anonymity.

Existing approaches to on-chain identity fall into four categories,
each with significant limitations:

\begin{enumerate}
\def\labelenumi{\arabic{enumi}.}
\item
  \textbf{Centralized attestation.} A trusted third party (e.g., KYC
  provider) verifies identity off-chain and issues an on-chain
  attestation. This centralizes trust, introduces a single point of
  failure, and leaks metadata revealing that a particular address was
  verified by a specific provider.
\item
  \textbf{Hardware-dependent verification.} Systems such as zkPassport
  \cite{ref2} require NFC readers to access passport chips, while Worldcoin
  \cite{ref3} depends on purpose-built biometric scanners (the Orb). These
  approaches limit accessibility to users with specific hardware.
\item
  \textbf{Direct credential submission.} Users submit identity documents
  to smart contracts or oracles, permanently recording personal data on
  an immutable ledger---a fundamental privacy violation.
\item
  \textbf{Decentralized Identifiers (DIDs).} W3C DID-based systems such
  as Polygon ID and Veramo require building entirely new credential
  issuance infrastructure---new issuers, new trust registries, and new
  verification workflows. While architecturally promising, DIDs face a
  cold-start problem: they cannot leverage the billions of credentials
  already issued by governments and CAs, and regulatory acceptance
  remains uncertain. Deployment timelines of 3--5 years for ecosystem
  bootstrapping limit their near-term applicability to
  compliance-sensitive domains.
\end{enumerate}

None of these approaches simultaneously achieves \textbf{verifiability}
(anyone can check that an address is backed by a valid credential),
\textbf{privacy} (no personal data is revealed),
\textbf{decentralization} (no single entity can forge or revoke
attestations), \textbf{accessibility} (no specialized hardware
required), and \textbf{immediate deployability} (no new issuance
infrastructure needed).

\subsection{Key Insight}\label{key-insight}

The dominant paradigm in blockchain identity research---Decentralized
Identifiers (DIDs)---attempts to build \emph{new} trust systems from
scratch. We argue for an orthogonal approach: \emph{bridging existing
trust} to the blockchain.

A vast, government-grade trust infrastructure already exists: the X.509
Public Key Infrastructure. Over 4 billion X.509 certificates are active
globally, issued by Certificate Authorities (CAs) for purposes ranging
from TLS to national identity. In Korea alone, approximately 20 million
NPKI certificates are actively used for banking, government services,
and e-commerce---each carrying legal weight under the Electronic
Signatures Act \cite{ref27}. These certificates embed RSA or ECDSA
signatures from trusted CAs, providing a cryptographic chain of trust
that can be verified computationally---and therefore inside a
zero-knowledge circuit. The core insight of zk-X509 is that these
existing credentials, already trusted by governments and institutions,
can serve as blockchain identity anchors \emph{today}, without waiting
years for new DID ecosystems to mature.

\subsection{Proposed Solution}\label{proposed-solution}

zk-X509 resolves the transparency-privacy tension by verifying X.509
certificate ownership entirely inside a zkVM. The system proves the
following properties without revealing any certificate contents:

\begin{enumerate}
\def\labelenumi{\arabic{enumi}.}
\tightlist
\item
  \textbf{Certificate Chain Validity.} The full chain from user
  certificate through intermediate CAs to a trusted root CA is verified,
  with each link's RSA or ECDSA signature checked cryptographically.
\item
  \textbf{Temporal Validity.} Every certificate in the chain is checked
  against the proof generation timestamp.
\item
  \textbf{Private Key Ownership.} The user proves possession of the
  private key corresponding to the certificate's public key.
\item
  \textbf{Revocation Status.} The CRL is parsed and its CA signature
  verified inside the zkVM, then the user's serial number is checked
  against the revoked list---providing trustless revocation checking.
\item
  \textbf{Registrant Binding.} The proof is cryptographically bound to
  the user's blockchain address, preventing proof theft via
  front-running.
\item
  \textbf{Nullifier Generation.} A deterministic, privacy-preserving
  identifier is derived from the certificate for Sybil resistance.
\end{enumerate}

Thirteen values are revealed publicly: a \textbf{nullifier}, a
\textbf{CA Merkle root} (proving membership in the whitelisted CA set
without revealing which CA), a \textbf{timestamp}, the
\textbf{registrant address}, a \textbf{wallet index}, the certificate's
\textbf{expiry time} (\texttt{notAfter}), a \textbf{chain ID} (EIP-155
\cite{ref28}), a \textbf{registry address} (cross-DApp unlinkability), a
\textbf{CRL Merkle root}, and four optional \textbf{selective disclosure
hashes} (country, organization, organizational unit, common name). These
are committed as public outputs and verified on-chain by a Solidity
smart contract.

\subsection{Global Applicability and Primary
Target}\label{global-applicability-and-primary-target}

zk-X509 is designed to work with \textbf{any X.509 certificate from any
CA worldwide}. The smart contract maintains a configurable whitelist of
trusted CA root hashes, enabling deployment-specific trust policies: a
Korean DeFi protocol may whitelist only Korean NPKI CAs, while a global
DAO may whitelist government CAs from multiple nations simultaneously.

\textbf{Primary validation target: Korean NPKI.} Our implementation is
validated against the Korean National Public Key Infrastructure (NPKI)
as a concrete case study. Korean digital certificates (Accredited Certificate) are
issued by authorized CAs such as the Korea Financial Telecommunications
and Clearings Institute (KFTC), employing a 3-level certificate
chain (Root CA → Intermediate CA → User Certificate) with RSA-2048 and
SHA-256 or SHA-1 signatures. Certificates and private keys are imported
into the OS keychain, where the prover application delegates signing to
the platform keychain API (macOS Security.framework, Windows CNG) without
direct access to the private key material.

\textbf{Multi-national deployment.} The architecture supports
simultaneous whitelisting of CAs from multiple jurisdictions. For
example, a single \texttt{IdentityRegistry} deployment could accept
certificates from Korean NPKI ($\sim$20M users), Estonian eID
($\sim$1.3M e-residents), German eID --- all operating under
the eIDAS regulation \cite{ref26} --- and corporate PKI systems---each user
proving identity under their national CA without any cross-border
credential issuance. The \texttt{caMerkleRoot} in the public values
attests that the certificate was issued by one of the whitelisted CAs
without revealing which one, preserving jurisdictional privacy.
Applications requiring jurisdiction-specific logic can request the user
to additionally disclose \texttt{countryHash} via selective disclosure
(Section 3.11).

\subsection{Contributions}\label{contributions}

This paper makes the following contributions:

\begin{itemize}
\tightlist
\item
  A \textbf{system architecture} for a complete ZK-based X.509
  verification pipeline supporting full certificate chain verification
  (RSA and ECDSA), trustless CRL checking, signature-based ownership
  (private key never enters zkVM), registrant binding, configurable
  multi-wallet registration, automatic identity expiry, selective
  attribute disclosure, CA-anonymous verification via Merkle tree, and
  self-service wallet migration.
\item
  A \textbf{working implementation} using the SP1 zkVM for
  zero-knowledge computation, with Solidity smart contracts for on-chain
  verification, on-chain CA list management with auto-computed Merkle
  roots, CRL Merkle root validation, configurable
  \texttt{maxWalletsPerCert} policy, chain ID and registry address
  binding, selective disclosure via bitmask, OS keychain integration,
  and a web-based frontend with NPKI auto-discovery.
\item
  A \textbf{formal security analysis} with game-based definitions under
  the Dolev-Yao adversary model, establishing eight properties ---
  unforgeability, unlinkability, cross-service unlinkability, cross-chain replay resistance, double-registration resistance,
  front-running immunity, CA-membership hiding, and non-transferability --- with
  explicit reductions to standard cryptographic assumptions (EUF-CMA,
  SHA-256 collision resistance, ZK soundness).
\item
  A \textbf{performance evaluation} demonstrating practical feasibility:
  $\sim$11.8M SP1 cycles for single-level ECDSA P-256
  verification ($\sim$17.4M for RSA-2048) and
  $\sim$300K gas for on-chain registration (Groth16).
\end{itemize}

\subsection{Paper Organization}\label{paper-organization}

Section 2 provides background on X.509, zkVMs, related work, and a
detailed comparison with DID-based approaches. Section 3 presents the
system architecture and formal protocol specification. Section 4 details
the implementation. Section 5 formalizes the security analysis with
game-based definitions. Section 6 compares with alternative approaches.
Section 7 discusses limitations and future work. Section 8 concludes.

\section{Background and Related
Work}\label{background-and-related-work}

\subsection{X.509 Certificates and Certificate
Chains}\label{x.509-certificates-and-certificate-chains}

X.509 is the ITU-T standard for public key certificates, defined in RFC
5280 \cite{ref4}. An X.509 certificate binds a public key to an identity
through a digital signature from a Certificate Authority. The
certificate structure (ASN.1 DER encoding) contains:

\begin{itemize}
\tightlist
\item
  \textbf{TBSCertificate} (To-Be-Signed): Subject name, issuer name,
  serial number, validity period, subject public key, and extensions.
\item
  \textbf{SignatureAlgorithm}: OID identifying the signature scheme
  (e.g., \texttt{sha256WithRSAEncryption}).
\item
  \textbf{SignatureValue}: The CA's digital signature over the
  DER-encoded TBSCertificate.
\end{itemize}

In practice, most PKI deployments use multi-level certificate chains. A
user certificate is signed by an intermediate CA, which is in turn
signed by a root CA. Verification requires traversing the entire chain,
verifying each signature and validity period. Korean NPKI uses a 3-level
hierarchy: KISA Root CA → Authorized CA (e.g., KFTC) → User
Certificate.

\subsection{Zero-Knowledge Proofs and
zkVMs}\label{zero-knowledge-proofs-and-zkvms}

A zero-knowledge proof allows a prover to convince a verifier that a
statement is true without revealing any information beyond the truth of
the statement \cite{ref5}. The random oracle model \cite{ref15} provides a
standard framework for analyzing hash-based constructions in such
proofs. Formally, a ZK proof system \((P, V)\) for a language \(L\)
satisfies three properties: \textbf{completeness} (honest provers
convince honest verifiers), \textbf{soundness} (no cheating prover can
convince on false statements), and \textbf{zero-knowledge} (the verifier
learns nothing beyond the statement's truth).

Modern zkVMs extend this to arbitrary computation: a prover executes a
program inside a virtual machine and produces a succinct proof that the
computation was performed correctly. The development of succinct
non-interactive arguments of knowledge (SNARKs) \cite{ref14} established the
foundation for this approach, with subsequent systems such as Groth16
\cite{ref7}, PLONK \cite{ref16}, Bulletproofs \cite{ref31}, and STARKs \cite{ref17}
offering different trade-offs between proof size, verification time, and
trusted setup requirements. Zerocash \cite{ref32} demonstrated the practical
application of SNARKs to privacy-preserving blockchain transactions.

\textbf{SP1} (Succinct Processor 1) is a RISC-V-based zkVM developed by
Succinct \cite{ref6}, alongside alternatives such as RISC Zero \cite{ref29}. It
compiles standard Rust to RISC-V instructions executed inside the zkVM,
enabling complex operations such as ASN.1 parsing and RSA verification.
SP1 provides precompiled accelerators for SHA-256 and RSA modular
exponentiation, and supports on-chain verification via Groth16 \cite{ref7}
or PLONK \cite{ref16} proof systems. Circuit-level DSLs such as Circom
\cite{ref30} offer an alternative approach but lack the general-purpose
programmability of zkVMs.

\subsection{RSA Verification in
Zero-Knowledge}\label{rsa-verification-in-zero-knowledge}

RSA signature verification---the dominant operation in X.509 certificate
validation---requires modular exponentiation with a 2048-bit modulus:
computing \(s^e \mod n\) where \(s\) is the signature, \(e\) is the
public exponent, and \(n\) is the modulus. Naive implementation in a ZK
circuit is prohibitively expensive due to the cost of big-integer
arithmetic. SP1 mitigates this through precompiled accelerators that
implement modular arithmetic natively in the proof system, reducing the
cycle count from tens of millions to approximately 5.5 million for
RSA-2048.

\subsection{Related Work}\label{related-work}

We survey existing approaches to privacy-preserving on-chain identity
and position zk-X509 relative to them.

\textbf{zkPassport} \cite{ref2} generates ZK proofs of passport and
national ID card data read via NFC. While similar in spirit to zk-X509,
it requires NFC hardware and is limited to NFC-readable travel documents
and eIDs. zk-X509 is purely software-based and works with any X.509
certificate.

\textbf{Worldcoin} \cite{ref3} uses iris biometric scanning with a
purpose-built open-source device (the Orb) to generate unique identity
proofs. The hardware dependency and biometric data collection raise
accessibility concerns that zk-X509 avoids entirely.

\textbf{Polygon ID and DID-based systems} \cite{ref13, ref33} use W3C Verifiable
Credentials (VCs) with ZK proofs. While providing a flexible credential
framework, DID systems face a fundamental bootstrapping problem: they
require new credential issuers, trust registries, and verification
schemas to be established before any identity verification can occur.
This ``build from scratch'' approach contrasts sharply with zk-X509's
``bridge the existing'' philosophy. Furthermore, DID revocation depends
on issuer-maintained registries---a centralized dependency---whereas
zk-X509 verifies CA-signed CRLs trustlessly inside the zkVM. Regulatory
acceptance of DID credentials remains unresolved in most jurisdictions,
while X.509 certificates carry established legal standing.

\textbf{Semaphore} \cite{ref8} enables anonymous group membership proofs but
provides no mechanism for certificate-based identity verification. It
solves a different problem: anonymous signaling within a pre-defined
group.

\textbf{zk-email} \cite{ref9} proves ownership of emails by verifying DKIM
signatures in ZK. This is the closest analog to zk-X509 in approach
(verifying existing cryptographic signatures in ZK), but is limited to
email and does not provide the government-grade trust level of PKI
certificates.

\textbf{Soulbound Tokens (SBTs)} \cite{ref10} propose non-transferable
tokens as identity primitives. However, SBTs require a trusted issuer
and provide no mechanism for privacy-preserving credential verification.

\textbf{Anonymous credential systems.} The foundational work of Chaum
\cite{ref19} introduced the concept of privacy-preserving transactions, and
Camenisch and Lysyanskaya \cite{ref18} formalized anonymous credential
schemes based on bilinear maps. These systems enable selective attribute
disclosure but require purpose-built credential issuance --- they cannot
leverage existing X.509 certificates. zk-X509 achieves comparable
privacy properties by applying ZK proofs to credentials that already
exist at scale.

\textbf{Proof of Personhood.} Ford \cite{ref20} and Borge et al.~\cite{ref21}
formalized the concept of proof of personhood as a prerequisite for
digital democracy. zk-X509 instantiates this concept using existing
government-issued certificates rather than novel pseudonym party
protocols or biometric approaches.

\begin{longtable}[]{@{}
  >{\raggedright\arraybackslash}p{(\linewidth - 12\tabcolsep) * \real{0.1127}}
  >{\raggedright\arraybackslash}p{(\linewidth - 12\tabcolsep) * \real{0.1549}}
  >{\raggedright\arraybackslash}p{(\linewidth - 12\tabcolsep) * \real{0.1408}}
  >{\raggedright\arraybackslash}p{(\linewidth - 12\tabcolsep) * \real{0.1831}}
  >{\raggedright\arraybackslash}p{(\linewidth - 12\tabcolsep) * \real{0.2113}}
  >{\raggedright\arraybackslash}p{(\linewidth - 12\tabcolsep) * \real{0.0704}}
  >{\raggedright\arraybackslash}p{(\linewidth - 12\tabcolsep) * \real{0.1268}}@{}}
\toprule\noalign{}
\begin{minipage}[b]{\linewidth}\raggedright
System
\end{minipage} & \begin{minipage}[b]{\linewidth}\raggedright
Credential
\end{minipage} & \begin{minipage}[b]{\linewidth}\raggedright
Hardware
\end{minipage} & \begin{minipage}[b]{\linewidth}\raggedright
Trust Model
\end{minipage} & \begin{minipage}[b]{\linewidth}\raggedright
Existing Infra
\end{minipage} & \begin{minipage}[b]{\linewidth}\raggedright
CRL
\end{minipage} & \begin{minipage}[b]{\linewidth}\raggedright
Privacy
\end{minipage} \\
\midrule\noalign{}
\endhead
\bottomrule\noalign{}
\endlastfoot
zkPassport \cite{ref2} & Passport/eID & NFC required & Government CA & Yes
(passports/eIDs) & N/A & Full ZK \\
Worldcoin \cite{ref3} & Biometric & Orb required & World Foundation & No
& N/A & Partial \\
DID/VC \cite{ref13} & W3C VC & None & New issuers & No (must build) &
Issuer-dependent & Varies \\
Semaphore \cite{ref8} & Group key & None & Group admin & No & N/A & Full
ZK \\
zk-email \cite{ref9} & Email DKIM & None & Email providers & Yes (DKIM) & No
& Full ZK \\
\textbf{zk-X509} & \textbf{X.509 cert} & \textbf{None} &
\textbf{Government CAs} & \textbf{Yes (billions)} & \textbf{Trustless
ZK} & \textbf{Full ZK} \\
\end{longtable}

zk-X509 is, to our knowledge, the first system to bring existing X.509
PKI certificates into the blockchain ecosystem using zero-knowledge
proofs, combining government-grade trust with full certificate chain
verification, revocation checking, full privacy, and no hardware
requirements.

\subsection{zk-X509 vs Decentralized Identifiers
(DIDs)}\label{zk-x509-vs-decentralized-identifiers-dids}

Decentralized Identifier (DID) frameworks \cite{ref13, ref33} represent the
dominant paradigm in blockchain identity research. While architecturally
elegant, DID-based systems differ fundamentally from zk-X509 in their
trust assumptions and deployment requirements. We highlight the key
distinctions:

\textbf{Infrastructure dependency.} DID systems require bootstrapping
entirely new infrastructure: credential issuers, trust registries,
verification schemas, and holder wallets. In contrast, zk-X509 leverages
X.509 PKI---an infrastructure already deployed at global scale with over
4 billion active certificates. In Korea alone, approximately 20 million
NPKI certificates are actively used, providing an immediate user base
without any new issuance required.

\textbf{Trust model.} DID trust is issuer-dependent: a verifier must
decide \emph{which} DID issuers to trust, creating a fragmented trust
landscape. zk-X509 inherits the established CA trust model, where
governments have already designated trusted CAs through legal frameworks
(e.g., Korea's Electronic Signatures Act \cite{ref27}). This eliminates the
``who trusts whom?'' bootstrapping problem.

\textbf{Revocation mechanism.} DID revocation depends on the issuer
maintaining and publishing revocation registries---a centralized
dependency within a supposedly decentralized system. zk-X509 performs
trustless CRL verification inside the zkVM: the CRL's CA signature is
cryptographically verified, ensuring revocation data cannot be forged or
suppressed.

\textbf{Regulatory compliance.} DID frameworks lack clear regulatory
standing in most jurisdictions. X.509 certificates, particularly
national PKI certificates, carry legal weight: Korean NPKI certificates
are legally binding under the Electronic Signatures Act \cite{ref27}. This
makes zk-X509 immediately applicable to compliance-sensitive domains
(banking, government services) where DID acceptance remains unresolved.

\textbf{Time to deployment.} DID ecosystems require 3--5 years for
credential issuance, trust registry establishment, and ecosystem
adoption. zk-X509 can be deployed within 3--6 months by whitelisting
existing CA root hashes---no new credential issuance is needed.

\begin{longtable}[]{@{}
  >{\raggedright\arraybackslash}p{(\linewidth - 4\tabcolsep) * \real{0.2157}}
  >{\raggedright\arraybackslash}p{(\linewidth - 4\tabcolsep) * \real{0.6078}}
  >{\raggedright\arraybackslash}p{(\linewidth - 4\tabcolsep) * \real{0.1765}}@{}}
\toprule\noalign{}
\begin{minipage}[b]{\linewidth}\raggedright
Criterion
\end{minipage} & \begin{minipage}[b]{\linewidth}\raggedright
DID (e.g., Polygon ID, Veramo)
\end{minipage} & \begin{minipage}[b]{\linewidth}\raggedright
zk-X509
\end{minipage} \\
\midrule\noalign{}
\endhead
\bottomrule\noalign{}
\endlastfoot
Existing infrastructure & Not leveraged; new issuers required &
Leverages billions of X.509 certs \\
Trust model & Issuer-dependent, fragmented & Government CAs, legally
established \\
Revocation & Issuer-maintained registries & Trustless CRL verification
in zkVM \\
Hardware requirement & None & None \\
Regulatory compliance & Unresolved in most jurisdictions & Legally
binding (e.g., Korea E-Sig Act) \\
Time to deployment & 3--5 years (ecosystem bootstrap) & 3--6 months
(whitelist existing CAs) \\
Trust establishment cost & High (new ecosystem) & Low (existing
government trust) \\
Privacy & ZK proofs (varies by system) & Full ZK (nullifier + CA Merkle
root only) \\
\end{longtable}

\textbf{Complementary roles.} DID and zk-X509 are not mutually
exclusive. DID excels at creating \emph{new} trust relationships in
domains where no prior credential infrastructure exists. zk-X509 excels
at bridging \emph{existing} government-grade trust to the blockchain. In
a mature ecosystem, a user might hold both: a DID for Web3-native
credentials and a zk-X509 registration for government-backed identity
verification.

\section{System Architecture}\label{system-architecture}

\subsection{Overview}\label{overview}

The zk-X509 system comprises four components arranged in a layered
architecture:

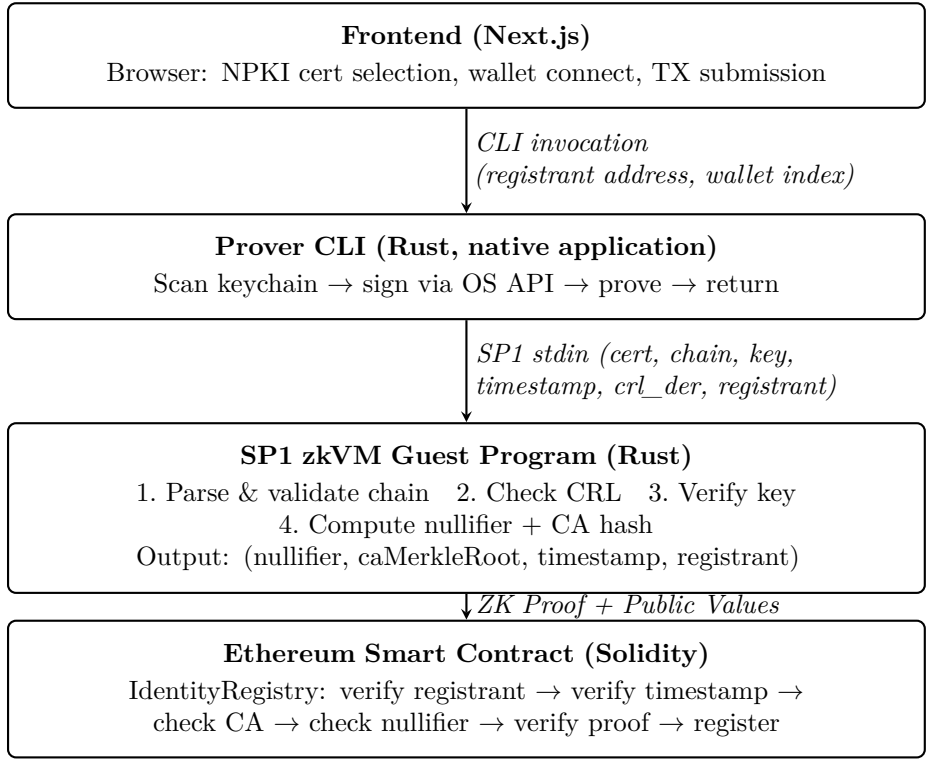
\begin{figure}[h]
\centering
\begin{tikzpicture}[
  box/.style={rectangle, draw, thick, rounded corners=3pt,
              text width=11.5cm, align=center, inner sep=8pt,
              font=\small},
  lbl/.style={font=\small\itshape, text width=6cm, align=left},
  arr/.style={->, thick, >=stealth}
]
\node[box] (fe) at (0,8.4) {
  \textbf{Frontend (Next.js)}\\[2pt]
  Browser: NPKI cert selection, wallet connect, TX submission
};
\node[box] (pr) at (0,5.6) {
  \textbf{Prover CLI (Rust, native application)}\\[2pt]
  Scan keychain $\rightarrow$ sign via OS API $\rightarrow$ prove $\rightarrow$ return
};
\node[box] (zk) at (0,2.4) {
  \textbf{SP1 zkVM Guest Program (Rust)}\\[2pt]
  1.~Parse \& validate chain\quad 2.~Check CRL\quad 3.~Verify key\\
  4.~Compute nullifier + CA hash\\
  Output: (nullifier, caMerkleRoot, timestamp, registrant)
};
\node[box] (sc) at (0,0) {
  \textbf{Ethereum Smart Contract (Solidity)}\\[2pt]
  IdentityRegistry: verify registrant $\rightarrow$ verify timestamp $\rightarrow$\\
  check CA $\rightarrow$ check nullifier $\rightarrow$ verify proof $\rightarrow$ register
};

\draw[arr] (fe) -- node[right, lbl]{CLI invocation\\(registrant address, wallet index)} (pr);
\draw[arr] (pr) -- node[right, lbl]{SP1 stdin (cert, chain, key,\\timestamp, crl\_der, registrant)} (zk);
\draw[arr] (zk) -- node[right, lbl]{ZK Proof + Public Values} (sc);
\end{tikzpicture}
\caption{zk-X509 system architecture overview.}
\label{fig:architecture}
\end{figure}

\subsection{Formal Protocol
Specification}\label{formal-protocol-specification}

We define the registration protocol formally. Let \(\mathcal{P}\) denote
the prover (user), \(\mathcal{S}\) the prover application (running locally
on \(\mathcal{P}\)'s machine), \(\mathcal{Z}\) the SP1 zkVM, and
\(\mathcal{V}\) the on-chain verifier (smart contract).

\textbf{Notation.} - \(\text{cert}\): DER-encoded user certificate -
\(\text{sk}\): User's private key (RSA PKCS\#1 DER or ECDSA SEC1 DER) -
\(\text{chain}\): Certificate chain
\([\text{cert}_{\text{inter}_1}, \ldots, \text{cert}_{\text{inter}_k}, \text{pk}_{\text{root}}]\)
where \(\text{pk}_{\text{root}}\) is the root CA's public key (SPKI DER)
- \(\text{CRL}\): DER-encoded Certificate Revocation List (signed by the
issuing CA) - \(\text{addr}\): \(\mathcal{P}\)'s Ethereum address (20
bytes) - \(t\): Current Unix timestamp - \(\mathcal{H}\): SHA-256

\textbf{Protocol.}

\begin{mdframed}[linewidth=0.5pt, innertopmargin=8pt, innerbottommargin=8pt]
\small
\textbf{Step 1.} $\mathcal{P} \rightarrow \mathcal{S}$:
\texttt{(cert\_index, addr, wallet\_index?, max\_wallets?, disclosure\_mask?)} via local API call. \texttt{cert\_index} identifies a certificate discovered by the keychain scanner; optional parameters have sensible defaults.

\medskip
\textbf{Step 2.} $\mathcal{S}$:
\begin{enumerate}[nosep,leftmargin=1.5em]
\item $(\text{cert}, \text{identity}) \leftarrow \text{Keychain.GetIdentity}(\text{cert\_index})$
\item $\text{CRL} \leftarrow \text{FetchCRL}(\text{cert.issuer})$ \hfill\textit{// from CA distribution point}
\item $\text{challenge} \leftarrow \mathcal{H}(\text{serial} \| \text{addr} \| \text{wallet\_index} \| t \| \text{chain\_id})$
\item $\text{ownership\_sig} \leftarrow \text{identity.Sign}(\text{challenge})$ \hfill\textit{// Security.framework}
\item $\text{nullifier\_sig} \leftarrow \text{identity.Sign}(\mathcal{H}(\text{``zk-X509-Nullifier-v2''}$\\
  $\| \text{registry\_addr} \| \text{chain\_id}))$
\end{enumerate}
Private key never leaves OS keychain --- no \texttt{Erase()} needed.

\medskip
\textbf{Step 3.} $\mathcal{S} \rightarrow \mathcal{Z}$:
$(\text{cert}, \text{ownership\_sig}, \text{nullifier\_sig}, \text{chain}, t, \text{CRL}, \text{addr},$\\
$\text{wallet\_index}, \text{max\_wallets}, \text{disclosure\_mask}, \text{ca\_merkle\_proof}, \text{ca\_merkle\_root})$\\
via SP1 stdin. \textbf{NOTE:} no private key.

\medskip
\textbf{Step 4.} $\mathcal{Z}$ (inside zkVM):
\begin{enumerate}[nosep,leftmargin=1.5em]
\item \textit{Parse and validate user certificate:}
  $\text{cert\_parsed} \leftarrow \text{ParseDER}(\text{cert})$;
  Assert $t \in [\text{notBefore}, \text{notAfter}]$
\item \textit{Verify certificate chain:}
  For $i = 0$ to $k{-}1$: parse $\text{inter}_i$, check temporal validity
\item \textit{Verify signatures} (RSA or ECDSA, auto-detected from OID):\\
  single-level: $\text{Sig.Verify}(\text{pk}_{\text{root}}, \text{cert.tbs}, \text{cert.sig})$;\\
  multi-level: verify each link $\text{cert} \rightarrow \text{inter}_0 \rightarrow \cdots \rightarrow \text{pk}_{\text{root}}$
\item \textit{Verify and check CRL} (if $\text{CRL} \neq \emptyset$):\\
  parse CRL, match issuer, check freshness ($\text{thisUpdate} \leq t \leq \text{nextUpdate}$),\\
  verify CA signature, assert serial $\notin$ revoked list
\item \textit{Verify key ownership:}\\
  $\text{challenge} \leftarrow \mathcal{H}(\text{serial} \| \text{addr} \| \text{wallet\_index} \| t \| \text{chain\_id})$;\\
  $\text{Sig.Verify}(\text{cert.pk}, \text{challenge}, \text{ownership\_sig})$
\item \textit{Verify wallet index:} Assert $\text{wallet\_index} < \text{max\_wallets}$
\item \textit{Verify nullifier signature:}\\
  $\text{Sig.Verify}(\text{cert.pk}, \mathcal{H}(\text{``zk-X509-Nullifier-v2''}$\\
  $\| \text{registry\_addr} \| \text{chain\_id}), \text{nullifier\_sig})$
\item \textit{Compute outputs:}
  $\text{nullifier} \leftarrow \mathcal{H}(\text{nullifier\_sig} \| \text{wallet\_index})$,
  $\text{caRootHash} \leftarrow \mathcal{H}(\text{pk}_{\text{root}})$,
  $\text{notAfter} \leftarrow \text{cert.notAfter}$
\item \textit{Verify CA Merkle membership:}
  $\text{MerkleVerify}(\text{caRootHash}, \pi_{\text{ca}}, \text{ca\_merkle\_root})$
\item \textit{Selective disclosure} (salted with private-key-derived salt):\\
  $\text{salt} \leftarrow \mathcal{H}(\text{``zk-X509-Disclosure-Salt-v1''} \| \text{nullifier\_sig})$;\\
  for each masked bit: $\text{hash} \leftarrow \mathcal{H}(\text{field} \| \text{salt})$ or $\texttt{0x0}$
\item \textit{Commit public values} (caMerkleRoot, NOT caRootHash)
\end{enumerate}

\medskip
\textbf{Step 5.} $\mathcal{Z} \rightarrow \mathcal{S}$:
$(\pi, \text{pubvals})$ where $\pi$ is the ZK proof.

\medskip
\textbf{Step 6.} $\mathcal{S} \rightarrow \mathcal{P}$:
$(\pi, \text{pubvals})$ returned to caller.

\medskip
\textbf{Step 7.} $\mathcal{P} \rightarrow \mathcal{V}$:
$\texttt{register}(\pi, \text{pubvals})$ via Ethereum transaction signed by addr.

\medskip
\textbf{Step 8.} $\mathcal{V}$ (on-chain verification):
\begin{enumerate}[nosep,leftmargin=1.5em]
\item Decode public values via \texttt{ABI.Decode(pubvals)}
\item Assert: $\text{registrant} = \texttt{msg.sender}$ \hfill\textit{// front-running}
\item Assert: $t_{\text{proof}} \leq \texttt{block.timestamp}$ \hfill\textit{// no future proofs}
\item Assert: $\texttt{block.timestamp} - t_{\text{proof}} \leq \texttt{maxProofAge}$ \hfill\textit{// freshness}
\item Assert: $\text{caMerkleRoot} = \text{contract.caMerkleRoot}$ \hfill\textit{// CA root match}
\item Assert: $\text{walletIndex} < \texttt{maxWalletsPerCert}$ \hfill\textit{// wallet limit}
\item Assert: $\text{notAfter} \geq \texttt{block.timestamp}$ \hfill\textit{// cert not expired}
\item Assert: $\text{chainId} = \texttt{block.chainid}$ \hfill\textit{// replay prevention}
\item Assert: $\text{registryAddress} = \texttt{address(this)}$ \hfill\textit{// cross-DApp binding}
\item If CRL enabled: assert $\text{crlMerkleRoot}$ match \hfill\textit{// CRL root}
\item $\texttt{SP1Verifier.verify}(\text{vkey}, \text{pubvals}, \pi)$ \hfill\textit{// ZK proof}
\item Assert: not revoked, not double-registered, expired or new
\item $\text{nullifierOwner}[\text{nullifier}] \leftarrow \texttt{msg.sender}$;
  $\text{verifiedUntil}[\texttt{msg.sender}] \leftarrow \text{notAfter}$
\item Emit \texttt{UserRegistered(msg.sender, nullifier)}
\end{enumerate}
\end{mdframed}

\subsection{Public Values
Structure}\label{public-values-structure}

The shared data structure between the ZK circuit and the smart contract
is:

\begin{table}[ht]
\centering
\small
\caption{\texttt{PublicValuesStruct} --- shared between ZK circuit and smart contract.}
\label{tab:public-values}
\begin{tabular}{@{}lll@{}}
\toprule
Field & Type & Description \\
\midrule
\texttt{nullifier} & \texttt{bytes32} & $\mathcal{H}(\text{nullifier\_sig} \| \text{walletIndex})$ \\
\texttt{caMerkleRoot} & \texttt{bytes32} & Merkle root of allowed CA set \\
\texttt{timestamp} & \texttt{uint64} & Proof generation timestamp \\
\texttt{registrant} & \texttt{address} & Wallet address bound to proof \\
\texttt{walletIndex} & \texttt{uint32} & Wallet slot index \\
\texttt{notAfter} & \texttt{uint64} & Certificate expiry \\
\texttt{chainId} & \texttt{uint64} & EIP-155 chain ID \\
\texttt{registryAddress} & \texttt{address} & Target registry address \\
\texttt{crlMerkleRoot} & \texttt{bytes32} & CRL Merkle root (\texttt{0} = disabled) \\
\texttt{countryHash} & \texttt{bytes32} & $\mathcal{H}(\text{country} \| \text{salt})$ or \texttt{0x0} \\
\texttt{orgHash} & \texttt{bytes32} & $\mathcal{H}(\text{organization} \| \text{salt})$ or \texttt{0x0} \\
\texttt{orgUnitHash} & \texttt{bytes32} & $\mathcal{H}(\text{org unit} \| \text{salt})$ or \texttt{0x0} \\
\texttt{commonNameHash} & \texttt{bytes32} & $\mathcal{H}(\text{common name} \| \text{salt})$ or \texttt{0x0} \\
\bottomrule
\end{tabular}
\end{table}

This struct is ABI-encoded using \texttt{alloy-sol-types} in Rust and
ABI-decoded in Solidity, ensuring binary compatibility across the stack.
The \texttt{caMerkleRoot} field replaces a direct \texttt{caRootHash}
with the Merkle root of the whitelisted CA set, hiding which specific CA
issued the certificate (Section 3.12). The \texttt{walletIndex} field
enables configurable multi-wallet registration (Section 3.6). The
\texttt{notAfter} field enables automatic identity expiry (Section 3.9).
The \texttt{chainId} and \texttt{registryAddress} fields ensure that
each blockchain and each contract deployment produces distinct
nullifiers. This design intentionally prevents cross-chain identity
portability, as different chains may have different trust requirements,
regulatory environments, and economic models. A user must register
separately on each chain, enabling per-chain and per-DApp Sybil
resistance policies. The \texttt{crlMerkleRoot} field commits the CRL
Sorted Merkle Tree root, enabling on-chain validation of revocation
checking (\texttt{bytes32(0)} disables CRL enforcement). The four
disclosure hash fields enable selective attribute disclosure (Section
3.11) --- each field is either the SHA-256 hash of the corresponding
certificate attribute salted with a private-key-derived salt (when
disclosed) or zero (when hidden), controlled by the user's
\texttt{disclosure\_mask}.

\subsection{ZK Guest Program}\label{zk-guest-program}

The guest program executes inside the SP1 zkVM and performs all
sensitive computations. A critical design principle is that \textbf{the
user's private key never enters the zkVM}. Instead, the prover application
uses the OS keychain to sign a challenge, and only the resulting
signature enters the circuit. This eliminates private key exposure from
the proving process entirely.

The program receives 23 inputs via SP1 stdin, organized into three
groups:

\textbf{Certificate \& ownership inputs (12):}

\begin{longtable}[]{@{}
  >{\raggedright\arraybackslash}p{(\linewidth - 6\tabcolsep) * \real{0.2121}}
  >{\raggedright\arraybackslash}p{(\linewidth - 6\tabcolsep) * \real{0.1818}}
  >{\raggedright\arraybackslash}p{(\linewidth - 6\tabcolsep) * \real{0.3333}}
  >{\raggedright\arraybackslash}p{(\linewidth - 6\tabcolsep) * \real{0.2727}}@{}}
\toprule\noalign{}
\begin{minipage}[b]{\linewidth}\raggedright
Input
\end{minipage} & \begin{minipage}[b]{\linewidth}\raggedright
Type
\end{minipage} & \begin{minipage}[b]{\linewidth}\raggedright
Visibility
\end{minipage} & \begin{minipage}[b]{\linewidth}\raggedright
Purpose
\end{minipage} \\
\midrule\noalign{}
\endhead
\bottomrule\noalign{}
\endlastfoot
\texttt{cert\_der} & \texttt{Vec\textless{}u8\textgreater{}} & Private &
DER-encoded user X.509 certificate \\
\texttt{ownership\_sig} & \texttt{Vec\textless{}u8\textgreater{}} &
Private & RSA or ECDSA signature over ownership challenge \\
\texttt{nullifier\_sig} & \texttt{Vec\textless{}u8\textgreater{}} &
Private & Deterministic signature of fixed domain string for
nullifier \\
\texttt{cert\_chain} &
\texttt{Vec\textless{}Vec\textless{}u8\textgreater{}\textgreater{}} &
Private & Chain:
\([\text{inter}_1, \ldots, \text{inter}_k, \text{pk}_{\text{root}}]\) \\
\texttt{current\_timestamp} & \texttt{u64} & Public (via output) & Unix
timestamp \\
\texttt{crl\_der} & \texttt{Vec\textless{}u8\textgreater{}} & Private &
DER-encoded CRL (empty = skip) \\
\texttt{registrant} & \texttt{{[}u8;\ 20{]}} & Public (via output) &
Wallet address \\
\texttt{wallet\_index} & \texttt{u32} & Public (via output) & Wallet
slot index (0-based) \\
\texttt{max\_wallets} & \texttt{u32} & Private & Max wallets per cert
(enforced in circuit) \\
\texttt{disclosure\_mask} & \texttt{u8} & Private & Bitmask: which cert
fields to reveal (bit 0=C, 1=O, 2=OU, 3=CN) \\
\texttt{ca\_merkle\_proof} &
\texttt{Vec\textless{}{[}u8;\ 32{]}\textgreater{}} & Private & Merkle
proof for CA membership (Section 3.12) \\
\texttt{ca\_merkle\_root} & \texttt{{[}u8;\ 32{]}} & Public (via output)
& Expected Merkle root of whitelisted CA set \\
\end{longtable}

\textbf{Domain separation inputs (2):}

\begin{longtable}[]{@{}
  >{\raggedright\arraybackslash}p{(\linewidth - 6\tabcolsep) * \real{0.2121}}
  >{\raggedright\arraybackslash}p{(\linewidth - 6\tabcolsep) * \real{0.1818}}
  >{\raggedright\arraybackslash}p{(\linewidth - 6\tabcolsep) * \real{0.3333}}
  >{\raggedright\arraybackslash}p{(\linewidth - 6\tabcolsep) * \real{0.2727}}@{}}
\toprule\noalign{}
\begin{minipage}[b]{\linewidth}\raggedright
Input
\end{minipage} & \begin{minipage}[b]{\linewidth}\raggedright
Type
\end{minipage} & \begin{minipage}[b]{\linewidth}\raggedright
Visibility
\end{minipage} & \begin{minipage}[b]{\linewidth}\raggedright
Purpose
\end{minipage} \\
\midrule\noalign{}
\endhead
\bottomrule\noalign{}
\endlastfoot
\texttt{registry\_address} & \texttt{{[}u8;\ 20{]}} & Public (via
output) & Target registry contract address (cross-DApp binding) \\
\texttt{chain\_id} & \texttt{u64} & Public (via output) & EIP-155 chain
ID (cross-chain binding) \\
\end{longtable}

\textbf{CRL Sorted Merkle Tree inputs (9):}

\begin{longtable}[]{@{}
  >{\raggedright\arraybackslash}p{(\linewidth - 6\tabcolsep) * \real{0.2121}}
  >{\raggedright\arraybackslash}p{(\linewidth - 6\tabcolsep) * \real{0.1818}}
  >{\raggedright\arraybackslash}p{(\linewidth - 6\tabcolsep) * \real{0.3333}}
  >{\raggedright\arraybackslash}p{(\linewidth - 6\tabcolsep) * \real{0.2727}}@{}}
\toprule\noalign{}
\begin{minipage}[b]{\linewidth}\raggedright
Input
\end{minipage} & \begin{minipage}[b]{\linewidth}\raggedright
Type
\end{minipage} & \begin{minipage}[b]{\linewidth}\raggedright
Visibility
\end{minipage} & \begin{minipage}[b]{\linewidth}\raggedright
Purpose
\end{minipage} \\
\midrule\noalign{}
\endhead
\bottomrule\noalign{}
\endlastfoot
\texttt{crl\_merkle\_root} & \texttt{{[}u8;\ 32{]}} & Public (via
output) & CRL Sorted Merkle Tree root (\texttt{{[}0;\ 32{]}} =
disabled) \\
\texttt{crl\_left\_leaf} & \texttt{{[}u8;\ 32{]}} & Private & Left
neighbor leaf in sorted tree \\
\texttt{crl\_right\_leaf} & \texttt{{[}u8;\ 32{]}} & Private & Right
neighbor leaf in sorted tree \\
\texttt{crl\_left\_proof} &
\texttt{Vec\textless{}{[}u8;\ 32{]}\textgreater{}} & Private & Merkle
proof for left neighbor \\
\texttt{crl\_left\_dirs} & \texttt{Vec\textless{}bool\textgreater{}} &
Private & Direction bits for left proof path \\
\texttt{crl\_right\_proof} &
\texttt{Vec\textless{}{[}u8;\ 32{]}\textgreater{}} & Private & Merkle
proof for right neighbor \\
\texttt{crl\_right\_dirs} & \texttt{Vec\textless{}bool\textgreater{}} &
Private & Direction bits for right proof path \\
\texttt{crl\_left\_index} & \texttt{u32} & Private & Index of left
neighbor in sorted tree \\
\texttt{crl\_right\_index} & \texttt{u32} & Private & Index of right
neighbor in sorted tree \\
\end{longtable}

The circuit asserts \texttt{wallet\_index\ \textless{}\ max\_wallets}
before proceeding. All private inputs remain hidden within the ZK proof.
The thirteen public values committed are: nullifier, caMerkleRoot,
timestamp, registrant, walletIndex, notAfter, chainId, registryAddress,
crlMerkleRoot, and four selective disclosure hashes (countryHash,
orgHash, orgUnitHash, commonNameHash --- zero when not disclosed).

\textbf{Certificate Parsing.} We use the \texttt{x509-parser} crate
(v0.16) with \texttt{default-features\ =\ false} to parse DER-encoded
certificates. Disabling default features avoids the \texttt{ring}
cryptography library, which contains platform-specific assembly
incompatible with the RISC-V zkVM target.

\textbf{Certificate Chain Verification.} The \texttt{cert\_chain} input
contains intermediate CA certificates followed by the root CA's public
key as the final element. The guest program verifies the signature
chain: user cert → intermediate CAs → root CA. For single-level PKI (no
intermediates), the chain contains only the root CA public key. For
Korean NPKI's 3-level hierarchy, the chain contains one intermediate CA
certificate and the root CA public key. Each intermediate certificate's
temporal validity is also checked.

\textbf{CA Signature Verification.} A unified
\texttt{verify\_cert\_signature()} function detects the signature
algorithm from the OID prefix and dispatches to the appropriate
verifier. RSA signatures (OID prefix \texttt{1.2.840.113549.1.1}) are
verified using the pure-Rust \texttt{rsa} crate (v0.9) with PKCS\#1 v1.5
padding and SHA-256, SHA-1, SHA-384, or SHA-512 digest selection. ECDSA
signatures (OID prefix \texttt{1.2.840.10045.4}) are verified using the
\texttt{p256} and \texttt{p384} crates, with the curve detected from the
signer's SPKI \texttt{namedCurve} OID (P-256 or P-384) independently of
the signature algorithm OID. This separation correctly handles cases
where the digest algorithm and curve are specified independently per RFC
5758.

\textbf{Trustless Certificate Revocation Checking.} The
\texttt{crl\_der} input contains a full DER-encoded Certificate
Revocation List. Unlike systems that rely on the host to provide
pre-filtered revocation data, zk-X509 performs \textbf{trustless CRL
verification} entirely inside the zkVM:

\begin{enumerate}
\def\labelenumi{\arabic{enumi}.}
\tightlist
\item
  \textbf{Parse} the DER-encoded CRL using
  \texttt{x509\_parser::revocation\_list}.
\item
  \textbf{Issuer matching}: Assert that the CRL's issuer matches the
  user certificate's issuer (serial numbers are issuer-scoped; checking
  against a CRL from a different issuer is meaningless).
\item
  \textbf{Freshness validation}: Assert
  \(\text{thisUpdate} \leq t \leq \text{nextUpdate}\), ensuring the CRL
  is current at the proof generation time.
\item
  \textbf{Signature verification}: Verify the CRL's signature (RSA or
  ECDSA, auto-detected) using the matching issuer's public key
  (intermediate CA for multi-level chains, root CA for single-level).
\item
  \textbf{Revocation check}: Assert that the user certificate's serial
  number is not in the CRL's revoked certificates list.
\end{enumerate}

This design ensures that a malicious host cannot supply a forged or
tampered CRL---the ZK proof cryptographically attests that the CRL was
signed by the legitimate issuing CA and was fresh at proof time. The CRL
data is not committed to public values; the proof attests only that
revocation was checked against a valid, CA-signed CRL.

\textbf{Signature-Based Key Ownership Verification.} Rather than
importing the private key into the zkVM, the prover application signs a
challenge using the OS keychain (macOS Secure Enclave, Windows TPM, or
software keystore). The challenge is
\(\mathcal{H}(\text{serial} \| \text{registrant} \| \text{wallet\_index} \| \text{timestamp} \| \text{chain\_id})\),
binding the ownership proof to the specific wallet, slot, proof
generation time, and chain. The ZK circuit verifies this signature using
the certificate's embedded public key:

\[\text{Sig.Verify}(\text{cert.pk}, \mathcal{H}(\text{serial} \| \text{registrant} \| \text{wallet\_index} \| \text{timestamp} \| \text{chain\_id}), \text{ownership\_sig})\]

The ownership verifier supports both RSA (PKCS\#1 v1.5 with SHA-256) and
ECDSA (P-256, P-384 with RFC 6979 \cite{ref22} deterministic nonces). The
key type is auto-detected from the certificate's SPKI algorithm OID: RSA
keys use direct \texttt{rsa} crate verification, while ECDSA keys use
the \texttt{p256}/\texttt{p384} crates with the curve determined from
the SPKI \texttt{namedCurve} parameter.

This approach has three advantages: (1) the private key never exists in
the prover's process memory---only the OS keychain handles it at
the hardware level, (2) the ownership proof is bound to the registrant
address and wallet index, preventing signature replay across wallets,
and (3) the timestamp binding prevents a compromised prover from
replaying a captured ownership signature in a later proof --- the
signature is only valid for the specific proof generation timestamp
committed as a public value.

\textbf{Nullifier Generation.} The nullifier is derived from a
deterministic signature rather than the certificate's public key:

\[\text{nullifier\_sig} = \text{Sign}(\text{sk}, \mathcal{H}(\text{"zk-X509-Nullifier-v2"} \| \text{contract\_address} \| \text{chain\_id}))\]
\[\text{nullifier} = \mathcal{H}(\text{nullifier\_sig} \| \text{wallet\_index})\]

The prover signs a fixed domain string with the certificate's private
key. RSA PKCS\#1 v1.5 and ECDSA with RFC 6979 \cite{ref22} deterministic
nonces are both inherently deterministic --- the same key always
produces the same signature, ensuring nullifier consistency. The ZK
circuit verifies the \texttt{nullifier\_sig} against the certificate's
public key before computing the nullifier.

This signature-based design prevents a critical linkability attack
present in public-key-based nullifiers. The certificate's public key is
semi-public data --- it is shared with banks, government portals, and
CRL distribution points during normal certificate usage. If the
nullifier were \(\mathcal{H}(\text{cert.pk} \| \text{wallet\_index})\),
any party possessing the certificate could compute all nullifiers and
track the user's on-chain registrations. With the signature-based
approach, only the private key holder can produce
\texttt{nullifier\_sig}, making the nullifier computationally
unpredictable without the private key. The \texttt{wallet\_index}
ensures that each wallet slot produces a distinct nullifier, enabling
configurable multi-wallet registration.

\subsection{Smart Contract}\label{smart-contract}

The \texttt{IdentityRegistry} contract manages on-chain state:

\begin{table}[ht]
\centering
\small
\caption{IdentityRegistry state variables.}
\label{tab:state-variables}
\begin{tabular}{@{}lll@{}}
\toprule
Variable & Type & Description \\
\midrule
\texttt{sp1Verifier} & \texttt{ISP1Verifier} (write-once) & On-chain proof verifier \\
\texttt{programVKey} & \texttt{bytes32} (write-once) & ZK program verification key \\
\texttt{maxWalletsPerCert} & \texttt{uint32} (write-once) & Max wallets per certificate \\
\texttt{caMerkleRoot} & \texttt{bytes32} & Merkle root of allowed CA set \\
\texttt{caLeaves} & \texttt{bytes32[]} & On-chain list of trusted CA hashes \\
\texttt{caExists} & \texttt{bytes32 $\Rightarrow$ bool} & Duplicate CA prevention \\
\texttt{crlMerkleRoot} & \texttt{bytes32} & CRL Merkle root (\texttt{0} = disabled) \\
\texttt{nullifierOwner} & \texttt{bytes32 $\Rightarrow$ address} & Nullifier $\rightarrow$ registered wallet \\
\texttt{revokedNullifiers} & \texttt{bytes32 $\Rightarrow$ bool} & Permanently revoked nullifiers \\
\texttt{verifiedUntil} & \texttt{address $\Rightarrow$ uint64} & Wallet $\rightarrow$ cert expiry timestamp \\
\texttt{owner} & \texttt{address} & Contract administrator \\
\texttt{pendingOwner} & \texttt{address} & For 2-step ownership transfer \\
\texttt{maxProofAge} & \texttt{uint256} & Max proof age (5 min--24 hours) \\
\texttt{paused} & \texttt{bool} & Emergency stop flag \\
\texttt{previousCaMerkleRoot} & \texttt{bytes32} & Previous CA Merkle root (grace) \\
\texttt{caMerkleRootUpdatedAt} & \texttt{uint256} & Timestamp of last CA root update \\
\texttt{caRootGracePeriod} & \texttt{uint256} & Grace period for old CA root \\
\texttt{minDisclosureMask} & \texttt{uint8} (write-once) & Minimum required disclosure fields \\
\bottomrule
\end{tabular}
\end{table}

The \texttt{maxWalletsPerCert} parameter is set once during proxy initialization, enabling
configurable registration policy per L2 deployment (see Section 3.6).
The \texttt{caLeaves} array stores the on-chain list of trusted CA
hashes, readable by anyone for off-chain Merkle proof generation. The
\texttt{caExists} mapping prevents duplicate CA additions. The
\texttt{crlMerkleRoot} stores the CRL Sorted Merkle Tree root for
on-chain CRL validation (\texttt{bytes32(0)} disables CRL checking). The
\texttt{nullifierOwner} mapping tracks which address owns each
nullifier, enabling \texttt{reRegister()}. The \texttt{verifiedUntil}
mapping stores the certificate's \texttt{notAfter} timestamp instead of
a boolean, enabling automatic identity expiry when the underlying
certificate expires (see Section 3.9).

\textbf{\texttt{register()}.} A shared \texttt{\_validateProof()}
function decodes public values and performs validation:

\begin{enumerate}
\def\labelenumi{\arabic{enumi}.}
\tightlist
\item
  \textbf{Registrant binding}: \texttt{registrant\ ==\ msg.sender} ---
  prevents front-running
\item
  \textbf{Timestamp freshness}:
  \texttt{block.timestamp\ -\ proofTimestamp\ ≤\ maxProofAge}
  (adjustable: 5 min to 24 hours, default 1 hour)
\item
  \textbf{CA Merkle root match}:
  \texttt{caMerkleRoot\ ==\ contract.caMerkleRoot} or
  \texttt{caMerkleRoot\ ==\ previousCaMerkleRoot} within
  \texttt{caRootGracePeriod} (default 24 hours) --- allows proofs
  generated before a CA list update to remain valid during the grace
  window
\item
  \textbf{Wallet index range}:
  \texttt{walletIndex\ \textless{}\ maxWalletsPerCert} --- enforces
  multi-wallet limit
\item
  \textbf{Certificate not expired}:
  \texttt{notAfter\ \textgreater{}=\ block.timestamp} --- rejects
  already-expired certificates
\item
  \textbf{Chain ID match}: \texttt{chainId\ ==\ block.chainid} ---
  prevents cross-chain replay
\item
  \textbf{Registry address match}:
  \texttt{registryAddress\ ==\ address(this)} --- prevents cross-DApp
  replay
\item
  \textbf{CRL root match} (if enabled):
  \texttt{crlMerkleRoot\ ==\ contract.crlMerkleRoot} --- validates
  revocation checking
\item
  \textbf{Proof validity}:
  \texttt{sp1Verifier.verifyProof(programVKey,\ publicValues,\ proof)}
\end{enumerate}

After validation, \texttt{register()} additionally checks:

\begin{enumerate}
\def\labelenumi{\arabic{enumi}.}
\setcounter{enumi}{9}
\tightlist
\item
  \textbf{Nullifier not revoked}:
  \texttt{revokedNullifiers{[}nullifier{]}\ ==\ false}
\item
  \textbf{Nullifier uniqueness}:
  \texttt{nullifierOwner{[}nullifier{]}\ ==\ address(0)}
\item
  \textbf{Address not already verified}:
  \texttt{verifiedUntil{[}msg.sender{]}\ \textless{}\ block.timestamp}
  --- allows re-registration after cert expiry
\item
  \textbf{State update}:
  \texttt{nullifierOwner{[}nullifier{]}\ =\ msg.sender;\ verifiedUntil{[}msg.sender{]}\ =\ notAfter}
\end{enumerate}

\textbf{\texttt{reRegister()}.} Enables self-service wallet migration
without admin approval. A user who loses access to their wallet can
generate a new proof with the same certificate and a new registrant
address. The contract verifies the proof, unverifies the old wallet, and
registers the new one. This eliminates the centralization concern of
admin-only revocation for wallet changes. The nullifier is reused (same
certificate, same wallet index), so the old wallet is automatically
displaced.

\textbf{Administrative functions:} -
\textbf{\texttt{addCA(bytes32\ caHash)}}: Adds a single trusted CA hash
to the on-chain list. Automatically recomputes \texttt{caMerkleRoot} via
\texttt{\_recomputeCaMerkleRoot()}. Reverts if the CA hash is zero or
already exists (duplicate prevention). -
\textbf{\texttt{addCAs(bytes32{[}{]}\ caHashes)}}: Batch adds multiple
trusted CA hashes in a single transaction. Recomputes the Merkle root
once at the end, saving gas compared to individual \texttt{addCA()}
calls. - \textbf{\texttt{removeCA(uint256\ index)}}: Removes a trusted
CA by index using swap-with-last-and-pop optimization. Automatically
recomputes \texttt{caMerkleRoot}. - \textbf{\texttt{getCaCount()} /
\texttt{getCaLeaves()}}: View functions returning the number of trusted
CAs and the full list of CA hashes, respectively. \texttt{getCaLeaves()}
enables off-chain users to compute Merkle proofs for proof generation. -
\textbf{\texttt{updateCaMerkleRoot(bytes32\ newRoot)}}: Manual override
for the CA Merkle root. Primarily used for initial setup or migration;
during normal operation, \texttt{addCA}/\texttt{removeCA} auto-compute
the root. - \textbf{\texttt{updateCrlMerkleRoot(bytes32\ newRoot)}}:
Updates the CRL Sorted Merkle Tree root. Set to \texttt{bytes32(0)} to
disable CRL checking on-chain. -
\textbf{\texttt{setMaxProofAge(uint256\ newAge)}}: Adjusts the maximum
allowed proof age, bounded between 5 minutes and 24 hours. Enables L2
deployments to tune the freshness window based on block time
characteristics. -
\textbf{\texttt{revokeIdentity(bytes32\ nullifier,\ bytes32\ reason)}}:
Permanently revokes a nullifier and unverifies the associated wallet.
This is irreversible --- the nullifier is added to
\texttt{revokedNullifiers} and can never be re-registered, even via
\texttt{reRegister()}. - \textbf{\texttt{pause()} / \texttt{unpause()}}:
Emergency stop mechanism to halt all registrations. -
\textbf{\texttt{transferOwnership(address)} /
\texttt{acceptOwnership()}}: Two-step ownership transfer preventing
accidental transfers.

\subsection{Configurable Registration
Policy}\label{configurable-registration-policy}

Different applications require different identity-to-wallet mappings.
DAO governance demands strict ``one person, one vote'' (1:1), while
decentralized exchanges need traders to verify multiple wallets for
trading, custody, and cold storage.

zk-X509 addresses this via the \texttt{maxWalletsPerCert} parameter, set
once at contract initialization:

\begin{longtable}[]{@{}
  >{\raggedright\arraybackslash}p{(\linewidth - 4\tabcolsep) * \real{0.3333}}
  >{\raggedright\arraybackslash}p{(\linewidth - 4\tabcolsep) * \real{0.2963}}
  >{\raggedright\arraybackslash}p{(\linewidth - 4\tabcolsep) * \real{0.3704}}@{}}
\toprule\noalign{}
\begin{minipage}[b]{\linewidth}\raggedright
Setting
\end{minipage} & \begin{minipage}[b]{\linewidth}\raggedright
Policy
\end{minipage} & \begin{minipage}[b]{\linewidth}\raggedright
Use Case
\end{minipage} \\
\midrule\noalign{}
\endhead
\bottomrule\noalign{}
\endlastfoot
\texttt{=\ 1} & Strict: one certificate, one wallet & DAO voting,
airdrops \\
\texttt{=\ 3} & Moderate: a few wallets per identity & DeFi (trading /
custody / cold) \\
\texttt{=\ N} & Flexible: many wallets, all verified & DEX,
multi-account platforms \\
\end{longtable}

The mechanism works through the \texttt{wallet\_index} parameter in the
nullifier:

\[\text{nullifier} = \mathcal{H}(\text{nullifier\_sig} \| \text{wallet\_index})\]

Each \texttt{wallet\_index} (0, 1, 2\ldots) produces a distinct
nullifier from the same deterministic signature. The ZK circuit enforces
\texttt{wallet\_index\ \textless{}\ max\_wallets}, and the smart
contract independently verifies
\texttt{walletIndex\ \textless{}\ maxWalletsPerCert}. Setting
\texttt{maxWalletsPerCert\ =\ 1} reduces to the strict 1:1
Sybil-resistant mode. Regardless of the setting, every verified wallet
is backed by a real, government-issued certificate.

This parameterization enables a single zk-X509 deployment on an L2 to
serve multiple protocols with different trust requirements.

\subsection{Self-Service
Re-Registration}\label{self-service-re-registration}

A critical limitation of naive nullifier-based systems is wallet
lock-in: if a user loses access to their wallet, the nullifier is
consumed and the certificate becomes permanently unusable. Traditional
solutions require admin intervention, introducing centralization.

zk-X509 solves this with \texttt{reRegister()}: a user generates a new
proof with the same certificate but a new registrant address. The
contract verifies the proof, unverifies the old wallet, and registers
the new one. No admin approval is required---the ZK proof itself serves
as authentication that the caller owns the certificate.

This design ensures that wallet migration is \textbf{self-sovereign}:
users control their own identity lifecycle without depending on any
centralized party.

\subsection{OS Keychain
Integration}\label{os-keychain-integration}

The prover application scans the OS keychain for identities (certificate +
private key pairs) via platform-native APIs. On macOS, the login
keychain is queried through Security.framework; on Windows, the
certificate store is accessed via CNG (Cryptography API: Next
Generation).

The private key is managed entirely by the OS keychain and \textbf{never
leaves the secure hardware boundary}. Signing is performed by calling
the keychain's signing API (\texttt{SecKey.createSignature()} on macOS)
--- the prover application receives only the resulting signature bytes and
never accesses the raw private key material. The signing algorithm (RSA
PKCS\#1 v1.5 or ECDSA) is auto-detected from the certificate's SPKI
algorithm OID. This model provides the strongest private key isolation:
even the prover's process memory never contains the key.

For each discovered identity, the scanner extracts metadata (subject,
issuer, serial number, expiry) for display in the frontend's certificate
selection UI. Users authenticate via the OS-level keychain prompt (e.g.,
macOS password dialog or Touch ID), after which the keychain generates
\texttt{ownership\_sig} and \texttt{nullifier\_sig} without exposing the
private key to any user-space process.

\subsection{Automatic Identity
Expiry}\label{automatic-identity-expiry}

A subtle but critical issue in on-chain identity systems is
\textbf{credential staleness}: once a wallet is marked as verified, it
typically remains so indefinitely, even after the underlying certificate
expires or is revoked. This creates a disconnect between the certificate
lifecycle and the on-chain state.

zk-X509 resolves this by committing the certificate's \texttt{notAfter}
timestamp as a public value. The smart contract stores this in
\texttt{verifiedUntil{[}address{]}} instead of a boolean flag. The
\texttt{isVerified()} function checks
\texttt{verifiedUntil{[}user{]}\ \textgreater{}=\ block.timestamp},
causing verification to automatically lapse when the certificate
expires. Users must re-prove with a renewed certificate to maintain
their verified status.

This design has two advantages: (1) on-chain identity tracks the
real-world credential lifecycle without manual intervention, and (2) it
creates a natural re-verification cycle that limits the damage window if
a certificate is compromised --- the compromised identity expires
automatically.

\subsection{Private Key Isolation}\label{private-key-isolation}

A critical design decision in zk-X509 is that the \textbf{private key
never enters the zkVM}. The circuit receives only two deterministic
signatures (\texttt{ownership\_sig}, \texttt{nullifier\_sig}) as inputs,
both generated on the user's local device via the OS keychain (macOS
Security.framework, Windows CNG). On devices with hardware-backed
keystores (Secure Enclave, TPM), the key may never leave secure
hardware. This architectural
separation of signing and proving ensures that the most sensitive
cryptographic material --- the private key --- remains strictly confined
to local hardware.

The proving flow is entirely local:

\begin{enumerate}
\def\labelenumi{\arabic{enumi}.}
\tightlist
\item
  \textbf{Signing ($\sim$1 second).} The prover application
  generates \texttt{ownership\_sig} and \texttt{nullifier\_sig} using
  the OS keychain. The private key never leaves the keychain --- only the resulting signature bytes are returned to the prover process.
\item
  \textbf{Proving ($\sim$5 minutes CPU).} The SP1 zkVM
  executes the guest program locally, receiving only the signatures,
  certificate, and chain data as inputs. The private key is never part
  of the zkVM witness.
\item
  \textbf{Submission.} The user submits the proof and public values to
  the blockchain via \texttt{register()}.
\end{enumerate}

\textbf{Comparison with other systems.} In zk-email, the raw email
content and DKIM signature enter the circuit as private inputs; the DKIM
signing key itself remains on the email provider's server and never
enters the circuit. In Semaphore, the user's secret identity enters the
circuit directly. If proof generation in either system were offloaded to
a third party, these private witness data would be exposed. In zk-X509,
the private key never leaves the user's device at any stage --- the zkVM
witness contains only signatures and publicly derivable certificate
data. This provides a stronger security model in the dimension of
private key exposure: no credential secret enters the ZK circuit,
whereas other systems require at least some private credential data as
witness inputs.

\subsection{Selective Attribute
Disclosure}\label{selective-attribute-disclosure}

Prior sections describe a binary identity model: the verifier learns
only ``this wallet holds a valid certificate'' without any attributes.
While this suffices for simple Sybil resistance, real-world applications
often require \textbf{granular attribute verification}: ``this user is
from country X'' or ``this user belongs to organization Y'' --- without
revealing other attributes like name or ID number.

zk-X509 implements selective disclosure via a \texttt{disclosure\_mask}
bitmask input to the ZK circuit:

\begin{longtable}[]{@{}llll@{}}
\toprule\noalign{}
Bit & Field & X.509 OID & Example \\
\midrule\noalign{}
\endhead
\bottomrule\noalign{}
\endlastfoot
0 & Country (C) & 2.5.4.6 & ``KR'', ``EE'', ``DE'' \\
1 & Organization (O) & 2.5.4.10 & ``KFTC'', ``Samsung'' \\
2 & Organizational Unit (OU) & 2.5.4.11 & ``Personal'', ``Engineering'' \\
3 & Common Name (CN) & 2.5.4.3 & (user's name --- typically hidden) \\
\end{longtable}

For each bit set in the mask, the circuit extracts the corresponding
field from the certificate's subject DN, hashes it with a
\textbf{private-key-derived salt}, and commits the hash as a public
value. For unset bits, zero is committed. The salt is computed as
\(\mathcal{H}(\text{"zk-X509-Disclosure-Salt-v1"} \| \text{nullifier\_sig})\),
where \texttt{nullifier\_sig} is the deterministic signature used for
nullifier generation. This salt is deterministic (same certificate
always produces the same salt) yet private (only the private key holder
can compute \texttt{nullifier\_sig}), preventing brute-force attacks on
small input spaces such as country codes ($\sim$200 values).
Without this salt, an attacker could precompute
\(\mathcal{H}(\text{"KR"})\), \(\mathcal{H}(\text{"US"})\), etc., and
match against on-chain \texttt{countryHash} values. With the salt, each
user's hashes are unique and unpredictable.

\textbf{User sovereignty.} The \texttt{disclosure\_mask} is chosen by
the user at proof generation time, not by the verifier. The same
certificate can produce different proofs for different applications: a
DAO voting contract may require only \texttt{countryHash}, while a
corporate DeFi protocol may additionally require \texttt{orgHash}. The
user decides what to reveal on a per-proof basis.

\textbf{Privacy guarantee.} Fields with mask bit = 0 produce a zero hash
in the public values, revealing no information. The ZK zero-knowledge
property ensures that even the \emph{existence} of undisclosed fields is
hidden --- the verifier cannot distinguish ``field is empty in the
certificate'' from ``field exists but was not disclosed.''

\subsection{CA-Anonymous Verification via Merkle
Tree}\label{ca-anonymous-verification-via-merkle-tree}

In a multi-national deployment, directly revealing \texttt{caRootHash}
(the SHA-256 hash of the root CA's public key) as a public value
discloses which CA issued the certificate --- effectively revealing the
user's jurisdiction (e.g., ``Korean CA'' vs ``Estonian CA''). This
narrows the anonymity set and may be unacceptable for privacy-sensitive
applications.

zk-X509 addresses this by replacing the direct \texttt{caRootHash}
output with a \textbf{Merkle membership proof}. The design works as
follows:

\begin{enumerate}
\def\labelenumi{\arabic{enumi}.}
\item
  \textbf{Off-chain setup.} The contract administrator constructs a
  Merkle tree whose leaves are the SHA-256 hashes of all whitelisted CA
  root public keys: \(\text{leaves} = \{h_1, h_2, \ldots, h_n\}\) where
  \(h_i = \mathcal{H}(\text{pk}_{\text{root}_i})\). The Merkle root
  \(M\) is stored on-chain.
\item
  \textbf{Proof generation.} The prover application computes
  \(\text{caRootHash} = \mathcal{H}(\text{pk}_{\text{root}})\) and
  generates a Merkle proof \(\pi_M\) demonstrating that
  \(\text{caRootHash}\) is a leaf in the tree with root \(M\). Both
  \(\pi_M\) and \(M\) are passed to the ZK circuit as private inputs.
\item
  \textbf{ZK verification.} Inside the zkVM, the circuit:

  \begin{itemize}
  \tightlist
  \item
    Computes
    \(\text{caRootHash} = \mathcal{H}(\text{pk}_{\text{root}})\)
    (already computed during chain verification)
  \item
    Verifies the Merkle proof:
    \(\text{MerkleVerify}(\text{caRootHash}, \pi_M, M)\)
  \item
    Commits \(M\) (the Merkle root) as the public value
    \texttt{caMerkleRoot}, instead of the leaf \(\text{caRootHash}\)
  \end{itemize}
\item
  \textbf{On-chain verification.} The smart contract stores a single
  \texttt{bytes32\ caMerkleRoot} and checks
  \(M = \text{contract.caMerkleRoot}\), confirming that the proof was
  generated against the current approved CA set.
\end{enumerate}

\textbf{Privacy improvement.} The on-chain disclosure is reduced from
``this user has a certificate from CA \(X\)'' to ``this user has a
certificate from \emph{one of} the \(n\) whitelisted CAs.'' The
anonymity set grows from 1 (a single CA's user base) to the union of all
whitelisted CAs' user bases. For a deployment whitelisting Korean NPKI
($\sim$20M), Estonian eID ($\sim$1.3M), and German eID
($\sim$46M), the anonymity set expands from a single
jurisdiction to $\sim$67M users.

\textbf{Quantitative anonymity analysis.} The effective anonymity set is
not simply the sum of all whitelisted CA user populations. An adversary
with auxiliary information can narrow the set through statistical
inference:

\begin{longtable}[]{@{}
  >{\raggedright\arraybackslash}p{(\linewidth - 4\tabcolsep) * \real{0.1818}}
  >{\raggedright\arraybackslash}p{(\linewidth - 4\tabcolsep) * \real{0.5455}}
  >{\raggedright\arraybackslash}p{(\linewidth - 4\tabcolsep) * \real{0.2727}}@{}}
\toprule\noalign{}
\begin{minipage}[b]{\linewidth}\raggedright
Factor
\end{minipage} & \begin{minipage}[b]{\linewidth}\raggedright
De-anonymization vector
\end{minipage} & \begin{minipage}[b]{\linewidth}\raggedright
Mitigation
\end{minipage} \\
\midrule\noalign{}
\endhead
\bottomrule\noalign{}
\endlastfoot
Registration timing & Correlate registration timestamp with CA issuance
patterns (e.g., Korean banking hours) & Randomized proof submission
delays \\
Gas payment source & Trace ETH funding of the registrant address to a
jurisdiction & Use fresh wallets funded via privacy-preserving
bridges \\
Selective disclosure & \texttt{countryHash} directly reveals
jurisdiction, reducing anonymity set to single-country users &
User-controlled --- do not disclose unless required \\
Transaction volume & Low-population CAs (e.g., Estonia
$\sim$1.3M) have fewer active blockchain users, making
correlation easier & Whitelist CAs with large user populations \\
Certificate expiry patterns & \texttt{notAfter} timestamp may correlate
with CA-specific validity periods (e.g., Korean NPKI: 1 year, Estonian
eID: 5 years) & Round \texttt{notAfter} to coarser granularity (future
work) \\
\end{longtable}

For a deployment with \(n\) whitelisted CAs, if CA \(i\) has \(N_i\)
active certificate holders, the theoretical anonymity set is
\(A = \sum_{i=1}^{n} N_i\). However, the \emph{effective} anonymity set
against an adversary who can estimate the prior probability
\(p_i = N_i / A\) of a user belonging to CA \(i\) is bounded by the
min-entropy: \(H_\infty = -\log_2(\max_i p_i)\). For the
Korean-Estonian-German deployment:
\(p_{\text{Germany}} = 46/67.3 \approx 0.68\), giving
\(H_\infty \approx 0.56\) bits --- meaning an adversary's best guess
(German eID) is correct 68\% of the time, even without selective
disclosure. Adding more CAs with comparable population sizes improves
the effective anonymity set. A deployment whitelisting 10+ CAs with
populations above 5M each would achieve \(H_\infty > 3\) bits, making
jurisdiction guessing impractical.

\textbf{Merkle tree construction.} A standard binary SHA-256 Merkle tree
\cite{ref25} with sorted-pair hashing is used:
\(H(\min(a,b) \| \max(a,b))\). Sorted-pair hashing prevents second
preimage attacks and eliminates the need for direction bits in the proof
path. For \(n\) whitelisted CAs, the proof consists of
\(\lceil \log_2 n \rceil\) hashes (e.g., 4 hashes for 16 CAs), adding
negligible overhead to the ZK circuit ($\sim${}\(\log_2 n\)
additional SHA-256 computations).

\textbf{CA set updates.} The contract maintains an on-chain list of
trusted CA hashes (\texttt{caLeaves{[}{]}}). When CAs are added or
removed via \texttt{addCA()}/\texttt{addCAs()}/\texttt{removeCA()}, the
contract automatically recomputes the Merkle root via
\texttt{\_recomputeCaMerkleRoot()} using the same sorted-pair SHA-256
algorithm as the zkVM. This ensures consistency between on-chain and
off-chain computations. Off-chain users call \texttt{getCaLeaves()} to
read the current CA set and compute their Merkle proofs. Proofs
generated against the old root will be rejected --- users must
regenerate proofs with the updated tree.

\section{Implementation}\label{implementation}

\subsection{Technology Stack}\label{technology-stack}

\begin{longtable}[]{@{}lll@{}}
\toprule\noalign{}
Component & Technology & Version \\
\midrule\noalign{}
\endhead
\bottomrule\noalign{}
\endlastfoot
ZK Prover & SP1 zkVM (Succinct) & v6.0.1 \\
ZK Guest Language & Rust (RISC-V target) & stable \\
Smart Contracts & Solidity + Foundry & \^{}0.8.20 / v1.5.1 \\
Prover CLI & Rust & --- \\
Frontend & Next.js + React + ethers.js & 16 / 19 / 6 \\
X.509 Parsing & x509-parser (no\_std) & 0.16 \\
RSA Verification & rsa (pure Rust) & 0.9 \\
ECDSA Verification & p256, p384 (pure Rust) & 0.13 \\
\end{longtable}

\subsection{Repository Structure}\label{repository-structure}

\begin{table}[ht]
\centering
\small
\caption{Repository structure.}
\label{tab:repo-structure}
\begin{tabular}{@{}ll@{}}
\toprule
Directory & Description \\
\midrule
\texttt{lib/} & Shared types (\texttt{PublicValuesStruct}) \\
\texttt{program/} & SP1 Guest program (zkVM) \\
\texttt{script/} & SP1 Host (CLI, keychain integration) \\
\texttt{contracts/} & Solidity (IdentityRegistry, tests, deploy scripts) \\
\texttt{backend/} & CA guide API server (Node.js/TypeScript) \\
\texttt{frontend/} & Next.js web UI \\
\texttt{certs/} & Test certificate generator (OpenSSL scripts) \\
\texttt{docs/} & Documentation and this paper \\
\bottomrule
\end{tabular}
\end{table}

\subsection{Performance Evaluation}\label{performance-evaluation}

\subsubsection{Off-Chain Cost: ZK
Proving}\label{off-chain-cost-zk-proving}

Measured on Apple M-series CPU using SP1 zkVM v6.0.1 execute mode with a
single-level certificate chain (user cert + root CA). The
signature-based nullifier design requires three signature verifications
per proof (ownership, nullifier, chain), making signature algorithm
selection the dominant cost factor:

\begin{longtable}[]{@{}
  >{\raggedright\arraybackslash}p{(\linewidth - 4\tabcolsep) * \real{0.4667}}
  >{\raggedleft\arraybackslash}p{(\linewidth - 4\tabcolsep) * \real{0.3667}}
  >{\centering\arraybackslash}p{(\linewidth - 4\tabcolsep) * \real{0.1667}}@{}}
\toprule\noalign{}
\begin{minipage}[b]{\linewidth}\raggedright
Configuration
\end{minipage} & \begin{minipage}[b]{\linewidth}\raggedleft
SP1 Cycles
\end{minipage} & \begin{minipage}[b]{\linewidth}\centering
vs RSA baseline
\end{minipage} \\
\midrule\noalign{}
\endhead
\bottomrule\noalign{}
\endlastfoot
\textbf{RSA-2048} (single-level, full disclosure) & 17,399,633 & --- \\
RSA-2048 (no disclosure, mask=0x00) & 17,384,766 & −0.1\% \\
RSA-2048 + CRL verification & 23,163,293 & +33.1\% \\
\textbf{ECDSA P-256} (single-level, full disclosure) & 11,803,639 &
−32.2\% \\
\textbf{ECDSA P-384} (single-level, full disclosure) & 47,775,211 &
+174.6\% \\
\end{longtable}

\textbf{Key findings:}

\begin{enumerate}
\def\labelenumi{\arabic{enumi}.}
\item
  \textbf{ECDSA P-256 is 32\% cheaper than RSA-2048} --- recommended for
  new certificate deployments. Single-level P-256 verification completes
  in 11.8M cycles.
\item
  \textbf{ECDSA P-384 is unexpectedly expensive} --- 2.7× more costly
  than RSA-2048. The 384-bit elliptic curve field operations cost
  approximately 4× more than 256-bit operations. P-384 should only be
  used when mandated by policy (e.g., CNSA Suite).
\item
  \textbf{CRL verification adds 33\%} --- approximately 5.8M additional
  cycles for a small test CRL (\textless1KB, 1 revoked entry).
  Real-world CRLs with thousands of entries would be significantly more
  expensive. For deployments where CRL cost is prohibitive, on-chain
  revocation via \texttt{revokeIdentity()} provides an alternative.
\item
  \textbf{Selective disclosure is essentially free} --- full disclosure
  (4 fields) vs none differs by only $\sim$15K cycles (0.1\%),
  as the SHA-256 cost is negligible compared to signature verification.
\end{enumerate}

\textbf{Cost breakdown (RSA single-level, estimated):}

\begin{longtable}[]{@{}lrc@{}}
\toprule\noalign{}
Operation & Estimated Cycles & Proportion \\
\midrule\noalign{}
\endhead
\bottomrule\noalign{}
\endlastfoot
RSA signature verify (ownership) & $\sim$5.7M & 33\% \\
RSA signature verify (chain) & $\sim$5.7M & 33\% \\
RSA signature verify (nullifier) & $\sim$5.7M & 33\% \\
SHA-256 hashing (all) & $\sim$200K & 1\% \\
Merkle proof verification & $\sim$40K & \textless1\% \\
Selective disclosure & $\sim$15K & \textless1\% \\
X.509 parsing + other & $\sim$100K & \textless1\% \\
\end{longtable}

Signature verification dominates at 99\% of total cycles. The primary
optimization lever is reducing signature count or switching to ECDSA
P-256.

\textbf{Multi-level chain cost.} Each additional chain level adds one
signature verification ($\sim$5.7M cycles for RSA-2048,
$\sim$3.9M for P-256):

\begin{longtable}[]{@{}lrr@{}}
\toprule\noalign{}
Chain Depth & RSA-2048 Cycles & P-256 Cycles \\
\midrule\noalign{}
\endhead
\bottomrule\noalign{}
\endlastfoot
1 (direct root signing) & $\sim$17.4M & $\sim$11.8M \\
2 (1 intermediate) & $\sim$23.1M & $\sim$15.7M \\
3 (2 intermediates) & $\sim$28.8M & $\sim$19.6M \\
\end{longtable}

\subsubsection{On-Chain Cost:
Verification}\label{on-chain-cost-verification}

Gas measurements on Ethereum (Foundry test environment):

\begin{longtable}[]{@{}
  >{\raggedright\arraybackslash}p{(\linewidth - 4\tabcolsep) * \real{0.4783}}
  >{\raggedright\arraybackslash}p{(\linewidth - 4\tabcolsep) * \real{0.2174}}
  >{\raggedright\arraybackslash}p{(\linewidth - 4\tabcolsep) * \real{0.3043}}@{}}
\toprule\noalign{}
\begin{minipage}[b]{\linewidth}\raggedright
Operation
\end{minipage} & \begin{minipage}[b]{\linewidth}\raggedright
Gas
\end{minipage} & \begin{minipage}[b]{\linewidth}\raggedright
Notes
\end{minipage} \\
\midrule\noalign{}
\endhead
\bottomrule\noalign{}
\endlastfoot
Contract deployment & $\sim$1,338,947 & IdentityRegistry +
SP1VerifierGroth16 \\
\texttt{register()} & $\sim$300,000 & With Groth16 on-chain
verifier \\
\texttt{addCA()} & 26,078 & Owner only, auto-recomputes Merkle root \\
\texttt{revokeIdentity()} & $\sim$8,500 & Owner only,
permanent \\
\texttt{isVerified()} & $\sim$2,600 & View function \\
\end{longtable}

The $\sim$300K gas cost for \texttt{register()} with Groth16
verification remains well within practical limits for Ethereum L1 and is
negligible on L2 rollups.

\subsubsection{End-to-End Latency}\label{end-to-end-latency}

\begin{longtable}[]{@{}
  >{\raggedright\arraybackslash}p{(\linewidth - 4\tabcolsep) * \real{0.3500}}
  >{\raggedright\arraybackslash}p{(\linewidth - 4\tabcolsep) * \real{0.3000}}
  >{\raggedright\arraybackslash}p{(\linewidth - 4\tabcolsep) * \real{0.3500}}@{}}
\toprule\noalign{}
\begin{minipage}[b]{\linewidth}\raggedright
Phase
\end{minipage} & \begin{minipage}[b]{\linewidth}\raggedright
Time
\end{minipage} & \begin{minipage}[b]{\linewidth}\raggedright
Notes
\end{minipage} \\
\midrule\noalign{}
\endhead
\bottomrule\noalign{}
\endlastfoot
Keychain signing (OS keychain) & \textless{} 1 second & Two signatures
(ownership + nullifier) \\
SP1 execute (no proof, CPU) & $\sim$15 seconds & Circuit
validation only, no proof \\
SP1 prove (CPU, Apple M1) & $\sim$5 minutes & Measured on Apple
M1, single-level P-256 \\
SP1 prove (GPU, estimated) & $\sim$1--2 minutes & GPU
acceleration via CUDA \\
On-chain verification & 1 block confirmation & Groth16 proof
verification \\
\end{longtable}

\textbf{Practical impact of proving time.} The $\sim$5 minute
CPU proving time must be evaluated in the context of actual usage
patterns. Identity registration is an infrequent operation: a user
generates a proof once per certificate per registry, and certificates
are typically valid for 1--3 years (Korean NPKI: 1 year, Estonian eID:
up to 5 years). The amortized proving cost is therefore $\sim$5
minutes per year --- comparable to the time required to complete a
traditional KYC onboarding process, which typically involves document
upload, video verification, and manual review over 1--3 days.

\begin{longtable}[]{@{}
  >{\raggedright\arraybackslash}p{(\linewidth - 4\tabcolsep) * \real{0.2941}}
  >{\raggedright\arraybackslash}p{(\linewidth - 4\tabcolsep) * \real{0.3235}}
  >{\raggedright\arraybackslash}p{(\linewidth - 4\tabcolsep) * \real{0.3824}}@{}}
\toprule\noalign{}
\begin{minipage}[b]{\linewidth}\raggedright
Scenario
\end{minipage} & \begin{minipage}[b]{\linewidth}\raggedright
Frequency
\end{minipage} & \begin{minipage}[b]{\linewidth}\raggedright
Proving time
\end{minipage} \\
\midrule\noalign{}
\endhead
\bottomrule\noalign{}
\endlastfoot
Initial registration & Once per registry & $\sim$5 minutes \\
Certificate renewal & Once per 1--3 years & $\sim$5 minutes \\
Wallet migration (\texttt{reRegister}) & Rare (wallet loss) &
$\sim$5 minutes \\
Verification check (\texttt{isVerified}) & Per transaction & 0 (on-chain
view call) \\
\end{longtable}

For user experience, proof generation can be performed in the background
while the user continues other tasks. The system does not require the
user to remain active during proving --- the process is non-interactive
after the initial keychain signing step.

\subsection{Testing}\label{testing}

The system includes three levels of testing:

\begin{enumerate}
\def\labelenumi{\arabic{enumi}.}
\tightlist
\item
  \textbf{Smart contract unit tests} (Foundry): Comprehensive test suites
  covering IdentityRegistry (registration, re-registration,
  double-registration prevention, registrant mismatch, CA Merkle root
  validation with grace period, timestamp validation, chain ID mismatch,
  registry address mismatch, wallet index out-of-range, certificate
  expiry, nullifier revocation, CA management, CRL Sorted Merkle root
  validation, pause/unpause, max proof age adjustment, minimum
  disclosure mask enforcement, and two-step ownership management) and
  RegistryFactory (registry creation, metadata tracking, beacon proxy
  deployment, and access control).
\item
  \textbf{SP1 execute mode}: Runs the ZK program without proof
  generation for fast iteration ($\sim$15 seconds), validating
  circuit logic.
\item
  \textbf{End-to-end integration}: Anvil local chain + contract
  deployment + prover CLI + frontend registration, verified with
  \texttt{cast} commands.
\end{enumerate}

\subsection{Supported Signature
Algorithms}\label{supported-signature-algorithms}

\begin{longtable}[]{@{}
  >{\raggedright\arraybackslash}p{(\linewidth - 4\tabcolsep) * \real{0.2083}}
  >{\raggedright\arraybackslash}p{(\linewidth - 4\tabcolsep) * \real{0.4583}}
  >{\raggedright\arraybackslash}p{(\linewidth - 4\tabcolsep) * \real{0.3333}}@{}}
\toprule\noalign{}
\begin{minipage}[b]{\linewidth}\raggedright
OID
\end{minipage} & \begin{minipage}[b]{\linewidth}\raggedright
Algorithm
\end{minipage} & \begin{minipage}[b]{\linewidth}\raggedright
Status
\end{minipage} \\
\midrule\noalign{}
\endhead
\bottomrule\noalign{}
\endlastfoot
1.2.840.113549.1.1.11 & sha256WithRSAEncryption & Supported \\
1.2.840.113549.1.1.5 & sha1WithRSAEncryption & Supported (legacy
NPKI) \\
1.2.840.113549.1.1.12 & sha384WithRSAEncryption & Supported \\
1.2.840.113549.1.1.13 & sha512WithRSAEncryption & Supported \\
1.2.840.10045.4.3.2 & ecdsa-with-SHA256 (P-256) & Supported \\
1.2.840.10045.4.3.3 & ecdsa-with-SHA384 (P-384) & Supported \\
\end{longtable}

SHA-1 support is included specifically for backward compatibility with
legacy Korean NPKI certificates that predate the SHA-256 migration.
ECDSA support (P-256 and P-384) extends compatibility to modern
certificate ecosystems that use elliptic curve cryptography, including
newer government PKI deployments and corporate CAs. The signature
algorithm OID determines the digest function, while the curve is
independently detected from the signer's SPKI \texttt{namedCurve} OID
\cite{ref23}, correctly handling the RFC 5758 separation of concerns.

\section{Security Analysis}\label{security-analysis}

\subsection{System Model}\label{system-model}

The system involves three entity types:

\begin{itemize}
\tightlist
\item
  \textbf{Prover} (\(\mathcal{P}\)): The user who owns an X.509
  certificate and wishes to register on-chain. \(\mathcal{P}\) runs the
  prover as a local application on their own machine.
\item
  \textbf{Verifier} (\(\mathcal{V}\)): The Ethereum smart contract
  (\texttt{IdentityRegistry}) that verifies proofs and manages
  registration state.
\item
  \textbf{Certificate Authority} (\(\text{CA}\)): A trusted authority
  (e.g., KFTC) whose root public key hash is whitelisted in
  \(\mathcal{V}\).
\end{itemize}

\textbf{Localhost assumption.} The prover runs as a native application on the user's machine. Private keys remain within the OS keychain and never enter general process memory; only signature results are passed to the proving engine. The security boundary is limited
to the user's own operating system, equivalent to the trust model of any
local application that reads certificate files (e.g., a web browser
using client certificates).

\subsection{Adversary Model}\label{adversary-model}

We adopt the \textbf{Dolev-Yao} adversary model \cite{ref11}. The adversary
\(\mathcal{A}\) has the following capabilities:

\begin{itemize}
\tightlist
\item
  \(\mathcal{A}\) can observe all transactions on the public blockchain,
  including proof bytes, public values, and transaction metadata (sender
  address, gas price, nonce).
\item
  \(\mathcal{A}\) can submit arbitrary transactions to the smart
  contract, including crafted proofs and replayed data from other users'
  transactions.
\item
  \(\mathcal{A}\) can monitor the mempool and attempt to front-run
  pending transactions by submitting competing transactions with higher
  gas prices.
\item
  \(\mathcal{A}\) can attempt to forge certificates or generate proofs
  with invalid inputs.
\end{itemize}

\textbf{Trust assumptions.} We assume:

\begin{itemize}
\tightlist
\item
  \textbf{A1 (Local security):} \(\mathcal{A}\) cannot compromise the
  user's local machine (i.e., cannot read files from the user's
  filesystem or inspect process memory).
\item
  \textbf{A2 (CA integrity):} The CA's private signing key has not been
  compromised.
\item
  \textbf{A3 (Cryptographic hardness):} RSA is secure under the
  factoring assumption \cite{ref12}; ECDSA is secure under the elliptic
  curve discrete logarithm assumption; SHA-256 is collision-resistant
  and preimage-resistant. Signature schemes satisfy EUF-CMA security
  \cite{ref24}.
\item
  \textbf{A4 (ZK soundness):} The SP1 proof system is computationally
  sound: no PPT adversary can generate a valid proof for a false
  statement with non-negligible probability.
\end{itemize}

\subsection{Security Definitions}\label{security-definitions}

We formalize eight security properties using game-based definitions. In
each game, \(\mathcal{A}\) interacts with a challenger \(\mathcal{C}\)
that simulates the system.

\paragraph{Definition 1
(Unforgeability)}\label{definition-1-unforgeability}

Consider the game \(\text{Exp}_{\mathcal{A}}^{\text{forge}}\):

\begin{mdframed}[linewidth=0.5pt, innertopmargin=6pt, innerbottommargin=6pt]
\small
\textbf{Game} $\text{Exp}_{\mathcal{A}}^{\text{forge}}$:
\begin{enumerate}[nosep,leftmargin=1.5em]
\item $\mathcal{C}$ deploys IdentityRegistry with verification key \textit{vkey} and whitelists CA root hashes $\{h_1, \ldots, h_n\}$
\item $\mathcal{A}$ is given: \textit{vkey}, $\{h_1, \ldots, h_n\}$, the contract address, and access to the public blockchain
\item $\mathcal{A}$ is \textbf{NOT} given: any valid certificate or private key
\item $\mathcal{A}$ outputs: $(\pi^*, \text{pubvals}^*)$
\item $\mathcal{A}$ wins if: $\mathcal{V}.\texttt{register}(\pi^*, \text{pubvals}^*)$ succeeds
\end{enumerate}
\end{mdframed}

\textbf{zk-X509 is unforgeable} if for all PPT adversaries
\(\mathcal{A}\):

\[\Pr[\text{Exp}_{\mathcal{A}}^{\text{forge}} = 1] \leq \text{negl}(\lambda)\]

\paragraph{Definition 2
(Unlinkability)}\label{definition-2-unlinkability}

Consider the game \(\text{Exp}_{\mathcal{A}}^{\text{link}}\):

\begin{mdframed}[linewidth=0.5pt, innertopmargin=6pt, innerbottommargin=6pt]
\small
\textbf{Game} $\text{Exp}_{\mathcal{A}}^{\text{link}}$:
\begin{enumerate}[nosep,leftmargin=1.5em]
\item $\mathcal{C}$ generates two valid certificates $(\text{cert}_0, sk_0)$ and $(\text{cert}_1, sk_1)$ both signed by the same CA
\item $\mathcal{C}$ generates registrations for both, producing nullifiers $n_0$ and $n_1$
\item $\mathcal{C}$ flips a random bit $b \in \{0, 1\}$
\item $\mathcal{A}$ is given: $n_b$, $n_{1-b}$ (in random order), and the caMerkleRoot
\item $\mathcal{A}$ is \textbf{NOT} given: the certificates, private keys, or serial numbers
\item $\mathcal{A}$ outputs: $b'$
\item $\mathcal{A}$ wins if: $b' = b$
\end{enumerate}
\end{mdframed}

\textbf{zk-X509 is unlinkable} if for all PPT adversaries
\(\mathcal{A}\):

\[\left| \Pr[\text{Exp}_{\mathcal{A}}^{\text{link}} = 1] - \frac{1}{2} \right| \leq \text{negl}(\lambda)\]

\paragraph{Definition 3 (Double-Registration
Resistance)}\label{definition-3-double-registration-resistance}

Consider the game \(\text{Exp}_{\mathcal{A}}^{\text{double}}\):

\begin{mdframed}[linewidth=0.5pt, innertopmargin=6pt, innerbottommargin=6pt]
\small
\textbf{Game} $\text{Exp}_{\mathcal{A}}^{\text{double}}$:
\begin{enumerate}[nosep,leftmargin=1.5em]
\item $\mathcal{C}$ deploys IdentityRegistry with $\texttt{maxWalletsPerCert} = 1$ and whitelists CA roots
\item $\mathcal{A}$ is given: one valid certificate $(\text{cert}, sk)$ and two addresses $\text{addr}_1$, $\text{addr}_2$
\item $\mathcal{A}$ outputs: two registration transactions $\text{tx}_1 = (\pi_1, \text{pubvals}_1)$ from $\text{addr}_1$ and $\text{tx}_2 = (\pi_2, \text{pubvals}_2)$ from $\text{addr}_2$, both using $\texttt{wallet\_index} = 0$
\item $\mathcal{A}$ wins if: both $\mathcal{V}.\texttt{register}(\text{tx}_1)$ and $\mathcal{V}.\texttt{register}(\text{tx}_2)$ succeed
\end{enumerate}
\end{mdframed}

\textbf{zk-X509 is double-registration resistant} if for all PPT
adversaries \(\mathcal{A}\):

\[\Pr[\text{Exp}_{\mathcal{A}}^{\text{double}} = 1] \leq \text{negl}(\lambda)\]

\paragraph{Definition 4 (Front-Running
Immunity)}\label{definition-4-front-running-immunity}

Consider the game \(\text{Exp}_{\mathcal{A}}^{\text{front}}\):

\begin{mdframed}[linewidth=0.5pt, innertopmargin=6pt, innerbottommargin=6pt]
\small
\textbf{Game} $\text{Exp}_{\mathcal{A}}^{\text{front}}$:
\begin{enumerate}[nosep,leftmargin=1.5em]
\item An honest user $\mathcal{P}$ generates a valid registration $\text{tx} = (\pi, \text{pubvals})$ for $\text{addr}_P$
\item $\mathcal{A}$ observes tx in the mempool before it is mined
\item $\mathcal{A}$ outputs: $\text{tx}' = (\pi', \text{pubvals}')$ from $\text{addr}_A \neq \text{addr}_P$ where $\mathcal{A}$ may copy, modify, or replay any data from tx
\item $\mathcal{A}$ wins if: $\mathcal{V}.\texttt{register}(\text{tx}')$ succeeds using any data derived from tx
\end{enumerate}
\end{mdframed}

\textbf{zk-X509 is front-running immune} if for all PPT adversaries
\(\mathcal{A}\):

\[\Pr[\text{Exp}_{\mathcal{A}}^{\text{front}} = 1] \leq \text{negl}(\lambda)\]

\paragraph{Definition 5 (CA-Membership
Hiding)}\label{definition-5-ca-membership-hiding}

Consider the game \(\text{Exp}_{\mathcal{A}}^{\text{ca-anon}}\):

\begin{mdframed}[linewidth=0.5pt, innertopmargin=6pt, innerbottommargin=6pt]
\small
\textbf{Game} $\text{Exp}_{\mathcal{A}}^{\text{ca-anon}}$:
\begin{enumerate}[nosep,leftmargin=1.5em]
\item $\mathcal{C}$ whitelists two CAs: $\text{CA}_0$ and $\text{CA}_1$, constructing a Merkle tree with roots $\{h_0 = \mathcal{H}(\text{pk}_0), h_1 = \mathcal{H}(\text{pk}_1)\}$ and Merkle root $M$
\item $\mathcal{C}$ generates a valid certificate $\text{cert}_b$ signed by $\text{CA}_b$, where $b \in \{0, 1\}$ is chosen uniformly at random
\item $\mathcal{C}$ generates a registration proof $\pi$ with public values pubvals (containing $\text{caMerkleRoot} = M$)
\item $\mathcal{A}$ is given: $\pi$, pubvals, $M$, and the public keys $\text{pk}_0$, $\text{pk}_1$
\item $\mathcal{A}$ is \textbf{NOT} given: $\text{cert}_b$, $sk_b$, or the Merkle proof path
\item $\mathcal{A}$ outputs: $b'$
\item $\mathcal{A}$ wins if: $b' = b$
\end{enumerate}
\end{mdframed}

\textbf{zk-X509 satisfies CA-membership hiding} if for all PPT
adversaries \(\mathcal{A}\):

\[\left| \Pr[\text{Exp}_{\mathcal{A}}^{\text{ca-anon}} = 1] - \frac{1}{2} \right| \leq \text{negl}(\lambda)\]

\paragraph{Definition 6
(Non-Transferability)}\label{definition-6-non-transferability}

Consider the game \(\text{Exp}_{\mathcal{A}}^{\text{transfer}}\):

\begin{mdframed}[linewidth=0.5pt, innertopmargin=6pt, innerbottommargin=6pt]
\small
\textbf{Game} $\text{Exp}_{\mathcal{A}}^{\text{transfer}}$:
\begin{enumerate}[nosep,leftmargin=1.5em]
\item $\mathcal{C}$ deploys IdentityRegistry and whitelists CA roots
\item An honest user $\mathcal{P}$ holds $(\text{cert}, sk)$ and does \textbf{NOT} cooperate with $\mathcal{A}$ (i.e., $\mathcal{P}$ does not sign any challenges, share signatures, or reveal any private data to $\mathcal{A}$)
\item $\mathcal{A}$ is given: access to the public blockchain, $\mathcal{P}$'s on-chain registration (if any), and $\mathcal{P}$'s certificate (public data)
\item $\mathcal{A}$ outputs: $(\pi^*, \text{pubvals}^*)$ with a registrant address controlled by $\mathcal{A}$
\item $\mathcal{A}$ wins if: $\mathcal{V}.\texttt{register}(\pi^*, \text{pubvals}^*)$ succeeds using $\mathcal{P}$'s certificate
\end{enumerate}
\end{mdframed}

\textbf{zk-X509 is non-transferable} if for all PPT adversaries
\(\mathcal{A}\):

\[\Pr[\text{Exp}_{\mathcal{A}}^{\text{transfer}} = 1] \leq \text{negl}(\lambda)\]

\textbf{Scope.} This definition captures \emph{involuntary} transfer
only. If the certificate holder voluntarily shares their private key or
pre-computed signatures, transfer becomes possible --- this is a
fundamental limitation shared by all credential systems (Section 5.5.5).

\subsection{Security Proofs}\label{security-proofs}

\paragraph{Theorem 1 (Unforgeability)}\label{theorem-1-unforgeability}

\emph{Under assumptions A2 (CA integrity), A3 (RSA hardness, SHA-256
collision resistance), and A4 (ZK soundness), zk-X509 satisfies
unforgeability (Definition 1).}

\textbf{Proof.} Suppose \(\mathcal{A}\) wins
\(\text{Exp}^{\text{forge}}\) with non-negligible probability. Then
\(\mathcal{A}\) produces \((\pi^*, \text{pubvals}^*)\) such that the
contract's \texttt{register()} succeeds. By assumption A4 (soundness),
the proof \(\pi^*\) attests that the ZK circuit executed correctly on
some witness
\((cert, \text{ownership\_sig}, chain, t, CRL, addr, \text{wallet\_index}, \text{max\_wallets})\).
The circuit verifies:

\begin{enumerate}
\def\labelenumi{(\alph{enumi})}
\item
  The certificate chain terminates at a root CA whose hash is a member
  of the whitelisted CA Merkle tree (verified via
  \texttt{caMerkleRoot}). Since \(\mathcal{A}\) does not possess a valid
  certificate signed by a whitelisted CA, \(\mathcal{A}\) must either
  forge the CA's signature --- RSA (contradicting A3 via the hardness of
  factoring \cite{ref12}) or ECDSA (contradicting A3 via the elliptic curve
  discrete logarithm assumption) --- or find a second preimage/collision
  in the Merkle tree to substitute a different CA (contradicting A3 via
  SHA-256 collision resistance).
\item
  The ownership signature verifies under the certificate's public key.
  Without the corresponding private key, \(\mathcal{A}\) cannot forge a
  valid signature --- whether RSA (contradicting A3 via factoring
  hardness) or ECDSA (contradicting A3 via ECDL hardness).
\end{enumerate}

In both cases, \(\mathcal{A}\)'s success contradicts one of the
assumptions. More precisely:

\[\Pr[\text{Exp}_{\mathcal{A}}^{\text{forge}} = 1] \leq \text{Adv}_{\mathcal{A}}^{\text{euf-cma}}(\text{CA.Sig}) + \text{Adv}_{\mathcal{A}}^{\text{euf-cma}}(\text{User.Sig}) + \text{Adv}_{\mathcal{A}}^{\text{col}}(\mathcal{H}) + \text{Adv}_{\mathcal{A}}^{\text{sound}}(\text{SP1}) \leq \text{negl}(\lambda)\]

where the four terms correspond to: (1) forging the CA's chain
signature, (2) forging the ownership signature, (3) finding a SHA-256
collision in the Merkle tree, and (4) breaking SP1 soundness.
\(\square\)

\paragraph{Theorem 2 (Unlinkability)}\label{theorem-2-unlinkability}

\emph{Under assumptions A3 (EUF-CMA security of the signature scheme,
SHA-256 collision resistance) and the zero-knowledge property of the SP1
proof system, zk-X509 satisfies unlinkability (Definition 2).}

\textbf{Proof.} The nullifier is
\(n = \mathcal{H}(\text{nullifier\_sig} \| \text{wallet\_index})\),
where
\(\text{nullifier\_sig} = \text{Sign}(\text{sk}, \mathcal{H}(\text{"zk-X509-Nullifier-v2"} \| \text{contract\_address} \| \text{chain\_id}))\).
The signature is computed using the certificate's private key, which is
known only to the certificate holder. The zero-knowledge property of the
proof system ensures that both the signature and the certificate
contents remain hidden.

To link a nullifier to a specific certificate, \(\mathcal{A}\) must
determine which \texttt{nullifier\_sig} was used. \(\mathcal{A}\) has
three strategies:

\begin{enumerate}
\def\labelenumi{(\alph{enumi})}
\item
  \textbf{Compute the signature directly.} \(\mathcal{A}\) possesses the
  certificate (which is semi-public data shared during normal
  certificate usage) and thus knows the public key. However, computing
  \texttt{nullifier\_sig} requires the private key. This reduces to the
  EUF-CMA security of the signature scheme:
  \(\text{Adv}^{\text{euf-cma}}(\text{Sig}) \leq \text{negl}(\lambda)\).
\item
  \textbf{Extract from the proof.} The ZK zero-knowledge property
  ensures that \(\pi\) reveals nothing about \texttt{nullifier\_sig}
  beyond what is already in the public values (which contain only
  \(\mathcal{H}(\text{nullifier\_sig} \| \text{wallet\_index})\), not
  \texttt{nullifier\_sig} itself).
\item
  \textbf{Invert the hash.} Recovering \texttt{nullifier\_sig} from
  \(n = \mathcal{H}(\text{nullifier\_sig} \| \text{wallet\_index})\)
  requires breaking preimage resistance of SHA-256.
\end{enumerate}

\(\mathcal{A}\)'s total advantage is bounded by:

\[\left| \Pr[b' = b] - \frac{1}{2} \right| \leq \text{Adv}_{\mathcal{A}}^{\text{euf-cma}}(\text{Sig}) + \text{Adv}_{\mathcal{A}}^{\text{zk}}(\text{SP1}) + \text{Adv}_{\mathcal{A}}^{\text{pre}}(\mathcal{H}) \leq \text{negl}(\lambda)\]

This is strictly stronger than a public-key-based nullifier
(\(\mathcal{H}(\text{cert.pk} \| \text{wallet\_index})\)), which would
be computable by anyone possessing the certificate. \(\square\)

\textbf{Caveat.} In the current implementation, \texttt{caMerkleRoot}
replaces the direct \texttt{caRootHash}, so on-chain observers learn
only that the certificate was issued by \emph{one of} the whitelisted
CAs --- the specific CA is hidden by the Merkle membership proof
(Section 3.12). This significantly enlarges the anonymity set in
multi-national deployments. Furthermore, the signature-based nullifier
ensures that even an adversary who independently obtains a user's
certificate (which contains the public key) cannot compute the nullifier
--- the private key is required to produce the deterministic
\texttt{nullifier\_sig}.

\paragraph{Theorem 3 (Cross-Service Unlinkability)}\label{theorem-3-cross-service-unlinkability} For any two
IdentityRegistry contracts \(C_1\) and \(C_2\) deployed at different
addresses, an adversary observing the on-chain nullifiers
\(\nu_1 \in C_1\) and \(\nu_2 \in C_2\) cannot determine whether
\(\nu_1\) and \(\nu_2\) were generated by the same user or by different
users, except with negligible advantage.

\emph{Under assumptions A3 (EUF-CMA security, SHA-256 preimage
resistance), A4 (ZK soundness and zero-knowledge), and the following
additional assumptions, zk-X509 satisfies cross-service unlinkability.}

\textbf{Additional assumptions.} We model \(\mathcal{H}\) (SHA-256) as a
random oracle. We assume the signature scheme provides strong
unforgeability (sEUF-CMA) and that the signature scheme is deterministic
(RSA PKCS\#1 v1.5 and ECDSA with RFC 6979), so distinct inputs produce
distinct signatures.

\textbf{Proof.} We proceed by a sequence of games.

\textbf{Game 0.} The real experiment. The challenger \(\mathcal{C}\)
generates two certificates \((\text{cert}_0, sk_0)\) and
\((\text{cert}_1, sk_1)\), and flips a coin \(b \in \{0, 1\}\).
\(\mathcal{C}\) registers user \(b\) in both \(C_1\) and \(C_2\), and
user \(1-b\) in neither. \(\mathcal{A}\) observes nullifiers
\(\nu_1 \in C_1\) and \(\nu_2 \in C_2\) (both from user \(b\)) plus
nullifiers from user \(1-b\) in unrelated registries, and outputs
\(b'\).

\textbf{Game 1.} Replace real ZK proofs with simulated proofs. By the
zero-knowledge property of SP1,
\(|\Pr_1[b' = b] - \Pr_0[b' = b]| \leq \text{Adv}_{\mathcal{A}}^{\text{zk}}(\text{SP1})\).
Now the proofs leak nothing about the witnesses.

\textbf{Game 2.} We show that \(\nu_1\) and \(\nu_2\) are
computationally independent even when generated by the same key. The
nullifier for registry \(C_i\) is:

\[\nu_i = \mathcal{H}(\text{Sign}(sk_b, \mathcal{H}(\text{"zk-X509-Nullifier-v2"} \| \text{addr}(C_i) \| \text{chain\_id}_i)) \| \text{wallet\_index})\]

Since \(\text{addr}(C_1) \neq \text{addr}(C_2)\), the domain inputs
\(d_1 \neq d_2\), producing distinct signatures
\(\sigma_1 = \text{Sign}(sk_b, d_1)\) and
\(\sigma_2 = \text{Sign}(sk_b, d_2)\). For \(\mathcal{A}\) to link
\(\nu_1\) to \(\nu_2\), \(\mathcal{A}\) must distinguish whether two
hash outputs \(\mathcal{H}(\sigma_1 \| \text{idx})\) and
\(\mathcal{H}(\sigma_2 \| \text{idx})\) share a common signing key. We
reduce this to two sub-problems:

\begin{enumerate}
\def\labelenumi{(\alph{enumi})}
\item
  \textbf{Recovering \(\sigma_i\) from \(\nu_i\).} Since the adversary
  observes two independent nullifiers
  \(\nu_1 = \mathcal{H}(\sigma_1 \| \text{idx})\) and
  \(\nu_2 = \mathcal{H}(\sigma_2 \| \text{idx})\), recovering either
  \(\sigma_i\) requires inverting the random oracle on the corresponding
  input. Each inversion contributes
  \(\text{Adv}_{\mathcal{A}}^{\text{pre}}(\mathcal{H})\), yielding a
  combined bound of
  \(2 \cdot \text{Adv}_{\mathcal{A}}^{\text{pre}}(\mathcal{H})\) for
  both nullifiers.
\item
  \textbf{Linking \(\sigma_1\) to \(\sigma_2\) without recovering them.}
  Even if \(\mathcal{A}\) could somehow relate the hash outputs without
  recovering the preimages, in the random oracle model
  \(\mathcal{H}(\sigma_1 \| \text{idx})\) and
  \(\mathcal{H}(\sigma_2 \| \text{idx})\) are independent random values
  as long as \(\sigma_1 \neq \sigma_2\). Since the signing inputs
  \(d_1 \neq d_2\) and the signature scheme is deterministic (RSA
  PKCS\#1 v1.5, ECDSA with RFC 6979), the signatures are distinct with
  overwhelming probability. In the random oracle model, the outputs
  reveal no information about any algebraic relationship between
  \(\sigma_1\) and \(\sigma_2\), rendering correlation impossible
  without preimage recovery.
\end{enumerate}

Combining all transitions:

\[\left| \Pr[b' = b] - \frac{1}{2} \right| \leq \text{Adv}_{\mathcal{A}}^{\text{zk}}(\text{SP1}) + 2 \cdot \text{Adv}_{\mathcal{A}}^{\text{pre}}(\mathcal{H}) \leq \text{negl}(\lambda)\]

where the factor of 2 accounts for the two independent hash inversions
required. The random oracle model ensures that hash outputs on distinct
inputs are independent, eliminating the need for a separate key-privacy
assumption. \(\square\)

\paragraph{Theorem 4 (Cross-Chain Replay Resistance)}\label{theorem-4-cross-chain-replay-resistance} A valid proof for
chain \(c_1\) cannot be accepted on chain \(c_2 \neq c_1\).

\emph{Under assumptions A1 (local security), A3 (EUF-CMA security), and
A4 (ZK soundness), zk-X509 satisfies cross-chain replay resistance.}

\textbf{Proof.} Suppose \(\mathcal{A}\) observes a valid registration
transaction \((\pi, \text{pubvals})\) on chain \(c_1\) and attempts to
replay or adapt it for acceptance on chain \(c_2\) where
\(c_2 \neq c_1\). We enumerate all strategies:

\begin{enumerate}
\def\labelenumi{(\alph{enumi})}
\item
  \textbf{Direct replay.} \(\mathcal{A}\) submits the identical
  \((\pi, \text{pubvals})\) on \(c_2\). The public values contain
  \(\text{chainId} = c_1\). The smart contract on \(c_2\) verifies
  \(\text{pubvals.chainId} == \text{block.chainid}\). Since
  \(\text{block.chainid} = c_2 \neq c_1\), the check fails
  deterministically. The proof \(\pi\) is never evaluated.
\item
  \textbf{Modify public values.} \(\mathcal{A}\) constructs
  \(\text{pubvals}'\) with \(\text{chainId} = c_2\) while reusing the
  original proof \(\pi\). The SP1 verifier checks
  \(\text{SP1Verifier.verifyProof}(\text{vkey}, \text{pubvals}', \pi)\).
  Since \(\pi\) was generated as a proof of correct execution with
  committed outputs \(\text{pubvals}\) (containing
  \(\text{chainId} = c_1\)), and
  \(\text{pubvals}' \neq \text{pubvals}\), the proof verification fails
  by the soundness of the SP1 proof system (A4). Formally:
\end{enumerate}

\[\Pr[\text{SP1Verifier.verify}(\text{vkey}, \text{pubvals}', \pi) = \text{true}] \leq \text{Adv}_{\mathcal{A}}^{\text{sound}}(\text{SP1}) \leq \text{negl}(\lambda)\]

\begin{enumerate}
\def\labelenumi{(\alph{enumi})}
\setcounter{enumi}{2}
\tightlist
\item
  \textbf{Generate a new proof for \(c_2\).} \(\mathcal{A}\) attempts to
  produce a fresh proof \(\pi'\) that commits \(\text{chainId} = c_2\).
  The ZK circuit requires a valid ownership signature over the challenge
  \(\mathcal{H}(\text{serial} \| \text{addr} \| \text{wallet\_index} \| t \| c_2)\).
  This challenge differs from the original (which used \(c_1\)). By A1,
  \(\mathcal{A}\) does not possess the user's private key \(sk\).
  Forging a valid signature on this new challenge reduces to breaking
  EUF-CMA security:
\end{enumerate}

\[\Pr[\text{forge ownership\_sig for } c_2] \leq \text{Adv}_{\mathcal{A}}^{\text{euf-cma}}(\text{Sig}) \leq \text{negl}(\lambda)\]

Similarly, the nullifier signature domain includes the chain ID:
\(\mathcal{H}(\text{"zk-X509-Nullifier-v2"} \| \text{registry\_addr} \| c_2)\).
Forging this signature adds a second EUF-CMA term.

Combining all strategies:

\[\Pr[\text{replay succeeds on } c_2] \leq \text{Adv}_{\mathcal{A}}^{\text{sound}}(\text{SP1}) + 2 \cdot \text{Adv}_{\mathcal{A}}^{\text{euf-cma}}(\text{Sig}) \leq \text{negl}(\lambda)\]

where the factor of 2 accounts for the two independent signature
forgeries (ownership and nullifier) required in strategy (c).
\(\square\)

\paragraph{Theorem 5 (Double-Registration
Resistance)}\label{theorem-3-double-registration-resistance}

\emph{Under assumption A4 (ZK soundness) and the determinism of the signature scheme (RSA PKCS\#1 v1.5, ECDSA with RFC 6979),
zk-X509 satisfies double-registration resistance (Definition 3).}

\textbf{Proof.} For a certificate with private key \(\text{sk}\) and
wallet index \(i\), the nullifier is deterministic:
\(n_i = \mathcal{H}(\text{Sign}(\text{sk}, \mathcal{H}(\text{"zk-X509-Nullifier-v2"} \| \text{contract\_address} \| \text{chain\_id})) \| i)\).
Since RSA PKCS\#1 v1.5 and ECDSA with RFC 6979 are deterministic
signature schemes, the same key always produces the same signature, and
thus the same nullifier. The ZK circuit enforces
\(i < \text{maxWalletsPerCert}\), limiting the number of distinct
nullifiers per certificate. After a registration with nullifier \(n_i\)
succeeds, the contract sets \texttt{nullifierOwner{[}n\_i{]}\ =\ addr}.
Any subsequent attempt to register the same nullifier fails because
\texttt{nullifierOwner{[}n\_i{]}\ !=\ address(0)}. The total number of
registrations per certificate is bounded by \texttt{maxWalletsPerCert}.
\(\square\)

\paragraph{Theorem 6 (Front-Running
Immunity)}\label{theorem-6-front-running-immunity}

\emph{Under assumptions A1 (local security), A3 (EUF-CMA security), and
A4 (ZK soundness), zk-X509 satisfies front-running immunity (Definition 4).}

\textbf{Proof.} The honest user's proof \(\pi\) commits
\texttt{registrant\ =\ addr\_P} as a public value. The contract verifies
\texttt{registrant\ ==\ msg.sender}. \(\mathcal{A}\) has two strategies:

\begin{enumerate}
\def\labelenumi{(\alph{enumi})}
\item
  \textbf{Replay the proof.} \(\mathcal{A}\) submits
  \((\pi, \text{pubvals})\) from \(\text{addr}_A\). Since
  \(\text{pubvals}\) contains \texttt{registrant\ =\ addr\_P} and
  \(\text{msg.sender} = \text{addr}_A \neq \text{addr}_P\), the
  registrant check fails.
\item
  \textbf{Modify pubvals.} \(\mathcal{A}\) changes \texttt{registrant}
  to \(\text{addr}_A\) in the public values. Since \texttt{pubvals} is
  an input to \texttt{SP1Verifier.verifyProof()}, altering it
  invalidates the proof verification (the proof was generated for the
  original public values).
\item
  \textbf{Generate a new proof.} \(\mathcal{A}\) would need \(sk\) to
  generate the required \texttt{ownership\_sig} and \texttt{nullifier\_sig}
  for a proof binding to \(\text{addr}_A\). By A1, \(\mathcal{A}\) does
  not have \(sk\).
\end{enumerate}

All strategies fail. \(\square\)

\paragraph{Theorem 7 (CA-Membership
Hiding)}\label{theorem-5-ca-membership-hiding}

\emph{Under assumption A4 (ZK soundness) and the zero-knowledge property
of the SP1 proof system, zk-X509 satisfies CA-membership hiding
(Definition 5). This property holds in the ideal ZK model; side-channel
attacks on the prover's execution environment (e.g., timing variations
correlated with certificate size) are outside the scope of this model
and apply equally to all ZK-based identity systems.}

\textbf{Proof.} The ZK circuit computes
\(\text{caRootHash} = \mathcal{H}(\text{pk}_{\text{root}})\) and
verifies a Merkle membership proof against the provided
\texttt{ca\_merkle\_root}. Only the Merkle root \(M\) is committed as a
public value; neither the leaf \(\text{caRootHash}\) nor the Merkle
proof path appears in the public outputs.

By the zero-knowledge property of SP1, the proof \(\pi\) reveals nothing
beyond the truth of the statement --- specifically, it does not reveal
which leaf was used. \(\mathcal{A}\) observes only \(M\) (the Merkle
root) and \(\pi\). Since \(M\) is identical regardless of which
\(\text{CA}_b\) issued the certificate, \(\mathcal{A}\)'s only strategy
is to extract information from \(\pi\). By the ZK property, \(\pi\) is
computationally indistinguishable from a simulated proof, so
\(\mathcal{A}\) gains no advantage:

\[\left| \Pr[b' = b] - \frac{1}{2} \right| \leq \text{Adv}_{\mathcal{A}}^{\text{zk}}(\text{SP1}) \leq \text{negl}(\lambda)\]

where \(\text{Adv}^{\text{zk}}\) is the advantage of distinguishing real
proofs from simulated ones. \(\square\)

\textbf{Remark.} If the deployment whitelists only a single CA (e.g.,
only Korean NPKI), CA-membership hiding is trivially satisfied (anonymity set =
1, but no information is revealed beyond what is already public). The
property becomes meaningful in multi-national deployments with
\(n \geq 2\) CAs.

\paragraph{Theorem 8
(Non-Transferability)}\label{theorem-8-non-transferability}

\emph{Under assumptions A1 (local security), A3 (cryptographic
hardness), and A4 (ZK soundness), zk-X509 satisfies non-transferability
(Definition 6).}

\textbf{Proof.} For \(\mathcal{A}\) to register using \(\mathcal{P}\)'s
certificate without \(\mathcal{P}\)'s cooperation, \(\mathcal{A}\) must
produce a valid proof \(\pi^*\) whose witness includes:

\begin{enumerate}
\def\labelenumi{(\alph{enumi})}
\item
  \textbf{An ownership signature} \(\text{ownership\_sig}\) that
  verifies under \(\mathcal{P}\)'s public key for a challenge binding
  \(\mathcal{A}\)'s address:
  \(\text{Sig.Verify}(\text{cert.pk}, \mathcal{H}(\text{serial} \| \text{addr}_A \| \text{wallet\_index} \| t \| \text{chain\_id}), \text{ownership\_sig})\).
  Without \(\mathcal{P}\)'s private key, forging this signature
  contradicts A3 (EUF-CMA security of RSA/ECDSA).
\item
  \textbf{A nullifier signature} \(\text{nullifier\_sig}\) that verifies
  under \(\mathcal{P}\)'s public key:
  \(\text{Sig.Verify}(\text{cert.pk},\)
  \(\mathcal{H}(\text{``zk-X509-Nullifier-v2''} \| \text{contract\_addr} \| \text{chain\_id}),\)
  \(\text{nullifier\_sig})\).
  Again, forging this without the private key contradicts A3.
\end{enumerate}

By A1, \(\mathcal{A}\) cannot extract these signatures from
\(\mathcal{P}\)'s machine. By A4, \(\mathcal{A}\) cannot produce a valid
proof with an invalid witness. Therefore \(\mathcal{A}\) cannot register
using \(\mathcal{P}\)'s certificate:

\[\Pr[\text{Exp}_{\mathcal{A}}^{\text{transfer}} = 1] \leq 2 \cdot \text{Adv}_{\mathcal{A}}^{\text{euf-cma}}(\text{Sig}) + \text{Adv}_{\mathcal{A}}^{\text{sound}}(\text{SP1}) \leq \text{negl}(\lambda)\]

where the factor of 2 accounts for the two independent signature
forgeries required (ownership and nullifier). \(\square\)

\subsection{Additional Attack
Analysis}\label{additional-attack-analysis}

\subsubsection{Timestamp Manipulation}\label{timestamp-manipulation}

\textbf{Attack.} The prover supplies a false timestamp to make an
expired certificate appear valid.

\textbf{Mitigation.} The timestamp is committed as a public value and
verified on-chain: - \texttt{proofTimestamp\ ≤\ block.timestamp}
(rejects future proofs) -
\texttt{block.timestamp\ -\ proofTimestamp\ ≤\ maxProofAge} (rejects
stale proofs; default 1 hour, adjustable 5 min--24 hours)

This bounds the manipulation window to \texttt{maxProofAge},
insufficient to exploit typical certificate validity periods of 1+
years. An adversary would need to advance the blockchain's clock, which
requires controlling block production---infeasible on Ethereum's
proof-of-stake consensus.

\subsubsection{CRL Integrity and
Freshness}\label{crl-integrity-and-freshness}

The CRL is verified trustlessly inside the zkVM: its signature (RSA or
ECDSA) is checked against the issuing CA's public key, and its temporal
validity (\(\text{thisUpdate} \leq t \leq \text{nextUpdate}\)) is
enforced. This prevents two attacks:

\begin{itemize}
\tightlist
\item
  \textbf{Forged CRL}: \(\mathcal{A}\) cannot supply a CRL not signed by
  the legitimate CA (signature verification inside zkVM).
\item
  \textbf{Stale CRL}: \(\mathcal{A}\) cannot supply an expired CRL
  (freshness check inside zkVM).
\end{itemize}

\textbf{Residual limitation.} The host selects \emph{which} valid CRL to
provide. If the CA has issued a newer CRL revoking the user's
certificate, a malicious host could still provide the older (but still
temporally valid) CRL that does not yet contain the revocation. This is
bounded by the CRL's validity window (typically 24--72 hours for Korean
NPKI). The CRL data is not committed to public values, so on-chain
consumers cannot independently verify which CRL was used. For stronger
guarantees, a CRL oracle (Section 7.2) could maintain an on-chain Merkle
root of revoked serials.

\textbf{Host-Provided but Cryptographically Authenticated CRL.} While
the CRL's cryptographic integrity is verified inside the zkVM (CA
signature and temporal validity), CRL \emph{freshness} ultimately
depends on the host providing the latest CRL from the CA's distribution
point. This creates an \textbf{Omission Attack} vector: a prover whose
certificate has been revoked can deliberately supply a stale---but still
temporally valid per \texttt{nextUpdate}---CRL that predates the
revocation entry. The maximum attack window equals the CRL update
period, which is typically 24 hours for most CAs (up to 72 hours for
some Korean NPKI CAs). During this window, a revoked certificate holder
can still generate valid proofs. For production deployments requiring
immediate revocation, we recommend two complementary mitigations: (1) an
\textbf{on-chain CRL oracle} (Section 7.2) that maintains a Merkle root
of revoked serials updated by a trusted operator or DAO, enabling the
circuit to commit the CRL Merkle root as a public value; and (2) the
existing \texttt{revokeIdentity()} admin function, which provides
immediate on-chain revocation independent of CRL propagation delays.

\subsubsection{Private Key Isolation}\label{private-key-isolation-1}

\textbf{Architecture.} The private key \textbf{never enters the zkVM or
the prover's general process memory}. The signature-based ownership
scheme (Section 3.4) delegates all private key operations to the OS
keychain:

\begin{enumerate}
\def\labelenumi{\arabic{enumi}.}
\tightlist
\item
  The prover application requests a signature from the OS keychain (macOS
  Security.framework, Windows CNG). The user authenticates via the OS-level prompt (e.g., password dialog or Touch ID).
\item
  The OS keychain signs the ownership challenge:
  \(\mathcal{H}(\text{serial} \| \text{registrant} \| \text{wallet\_index}\)
  \(\| \text{timestamp} \| \text{chain\_id})\).
\item
  Only the resulting \textbf{signature bytes} are returned to the prover process and passed to the SP1 zkVM as input.
\item
  The private key never leaves the OS keychain and never enters general process memory.
\end{enumerate}

\textbf{Security properties:} - The private key never appears in any
HTTP request or response. - The private key never enters the SP1 RISC-V
virtual machine. - On devices with hardware-backed keystores (macOS
Secure Enclave, Windows TPM), the private key may never exist in general
process memory at all --- the signing operation occurs within the secure
hardware. No \texttt{Debug} derive on key-holding structs; API access restricted to authenticated requests.

This provides a stronger security model in the dimension of private key
exposure compared to approaches that import the private key directly
into the ZK circuit, and exceeds the trust model of standard
certificate-using software (e.g., web browsers performing TLS client
authentication).

\subsubsection{Smart Contract Security}\label{smart-contract-security}

\begin{itemize}
\tightlist
\item
  \textbf{Reentrancy.} The \texttt{register()} function performs all
  validation checks and the external \texttt{verifyProof()} call before
  updating state. While the state updates occur after the external call,
  \texttt{verifyProof} is a pure verification function that either
  returns successfully or reverts---it has no callback mechanism or
  state-modifying side effects. The verifier contract
  (\texttt{ISP1Verifier}) is immutably set at deployment, preventing
  substitution with a malicious contract.
\item
  \textbf{Access control.} Administrative functions (\texttt{addCA},
  \texttt{addCAs}, \texttt{removeCA}, \texttt{updateCaMerkleRoot},
  \texttt{updateCrlMerkleRoot}, \texttt{revokeIdentity},
  \texttt{setMaxProofAge}, \texttt{pause}, \texttt{unpause}) are
  protected by the \texttt{onlyOwner} modifier. Ownership transfer uses
  a two-step pattern (\texttt{transferOwnership} →
  \texttt{acceptOwnership}) to prevent accidental transfers.
\item
  \textbf{Emergency stop.} The \texttt{pause()} function halts all
  registrations, providing an escape hatch if a critical vulnerability
  is discovered.
\item
  \textbf{Integer overflow.} Solidity \^{}0.8.x provides built-in
  overflow/underflow checks.
\end{itemize}

\subsubsection{Voluntary Credential
Sharing}\label{voluntary-credential-sharing}

\textbf{Attack.} A certificate holder voluntarily shares their private
key or pre-computed signatures (\texttt{ownership\_sig},
\texttt{nullifier\_sig}) with a third party, enabling the third party to
generate valid proofs and register under the original certificate.

\textbf{Scope.} This is a fundamental limitation shared by all
credential-based identity systems, including DID/VC, zkPassport, and
even biometric systems (where liveness detection can be circumvented).
No cryptographic mechanism can prevent a willing user from delegating
their credential usage.

\textbf{Analysis of the zk-X509 attack surface.} Compared to other
systems, zk-X509 offers partial mitigation:

\begin{enumerate}
\def\labelenumi{\arabic{enumi}.}
\item
  \textbf{Signature sharing (limited delegation).} If the user shares
  only \texttt{ownership\_sig} and \texttt{nullifier\_sig} (not the
  private key), the third party can register once but cannot generate
  new proofs for different registries, timestamps, or wallet indices.
  The delegation is scoped to the specific parameters embedded in the
  signatures.
\item
  \textbf{Private key sharing (full delegation).} If the user shares the
  private key itself, the third party gains full control. However, NPKI
  private keys are tied to real-world identity with legal consequences
  --- sharing a Korean NPKI certificate constitutes a criminal offense
  under the Electronic Signatures Act, creating a legal deterrent absent
  in pseudonymous systems.
\item
  \textbf{Economic disincentives.} In Sybil-resistant contexts (e.g.,
  DAO voting), the economic value of one additional identity is bounded
  by the per-identity reward. If the cost of obtaining a fraudulent
  certificate (legal risk, social engineering) exceeds this value,
  rational actors will not engage in credential sharing.
\item
  \textbf{Comparison with other systems.} Worldcoin's biometric approach
  theoretically prevents sharing (one cannot share an iris), but faces
  practical circumvention via coerced enrollment. DID systems face the
  same voluntary sharing problem as zk-X509. Certificate-based systems
  with legal backing (X.509, eID) arguably have stronger deterrents than
  purely cryptographic credentials due to the legal liability attached
  to the credential.
\end{enumerate}

\textbf{Open problem.} Fully preventing voluntary credential sharing
while maintaining privacy remains an open problem in the identity
literature. Hardware-bound credentials (e.g., certificates stored in
smart cards with non-exportable keys) offer a partial solution at the
cost of hardware dependency --- precisely the constraint zk-X509 is
designed to avoid.

\subsection{Privacy Properties
Summary}\label{privacy-properties-summary}

\begin{longtable}[]{@{}
  >{\raggedright\arraybackslash}p{(\linewidth - 4\tabcolsep) * \real{0.3448}}
  >{\raggedright\arraybackslash}p{(\linewidth - 4\tabcolsep) * \real{0.2759}}
  >{\raggedright\arraybackslash}p{(\linewidth - 4\tabcolsep) * \real{0.3793}}@{}}
\toprule\noalign{}
\begin{minipage}[b]{\linewidth}\raggedright
Property
\end{minipage} & \begin{minipage}[b]{\linewidth}\raggedright
Status
\end{minipage} & \begin{minipage}[b]{\linewidth}\raggedright
Guarantee
\end{minipage} \\
\midrule\noalign{}
\endhead
\bottomrule\noalign{}
\endlastfoot
Certificate subject (name, ID) & Hidden & ZK zero-knowledge property \\
Certificate serial number & Hidden & Not used in nullifier; hidden by
ZK \\
Certificate public key & Not linkable to nullifier & Signature-based
nullifier requires private key \\
Certificate attributes (C, O, OU, CN) & User-controlled & Disclosed only
if user sets disclosure\_mask bit \\
Identity expiry & Automatic & notAfter committed; verifiedUntil expires
on-chain \\
Private key & Never enters zkVM & Signature-based ownership; OS keychain
isolation \\
CA identity & Hidden & caMerkleRoot hides which CA; Merkle membership
proof (Section 3.12) \\
Wallet-to-certificate link & Unlinkable & Theorem 2 \\
Proof-to-address binding & Enforced & Theorem 6 \\
Double registration & Prevented & Theorem 5 \\
Non-transferability & Enforced (without cooperation) & Theorem 8;
voluntary delegation is a universal credential limitation \\
Multiple certs per wallet & Prevented & \texttt{verifiedUntil}
mapping \\
\end{longtable}

\section{Comparison with Alternative
Approaches}\label{comparison-with-alternative-approaches}

\begin{longtable}[]{@{}
  >{\raggedright\arraybackslash}p{(\linewidth - 12\tabcolsep) * \real{0.1429}}
  >{\raggedright\arraybackslash}p{(\linewidth - 12\tabcolsep) * \real{0.1169}}
  >{\raggedright\arraybackslash}p{(\linewidth - 12\tabcolsep) * \real{0.1558}}
  >{\raggedright\arraybackslash}p{(\linewidth - 12\tabcolsep) * \real{0.0909}}
  >{\raggedright\arraybackslash}p{(\linewidth - 12\tabcolsep) * \real{0.1299}}
  >{\raggedright\arraybackslash}p{(\linewidth - 12\tabcolsep) * \real{0.1948}}
  >{\raggedright\arraybackslash}p{(\linewidth - 12\tabcolsep) * \real{0.1688}}@{}}
\toprule\noalign{}
\begin{minipage}[b]{\linewidth}\raggedright
Criterion
\end{minipage} & \begin{minipage}[b]{\linewidth}\raggedright
zk-X509
\end{minipage} & \begin{minipage}[b]{\linewidth}\raggedright
DID/VC \cite{ref13}
\end{minipage} & \begin{minipage}[b]{\linewidth}\raggedright
zkKYC
\end{minipage} & \begin{minipage}[b]{\linewidth}\raggedright
SBT \cite{ref10}
\end{minipage} & \begin{minipage}[b]{\linewidth}\raggedright
zkPassport \cite{ref2}
\end{minipage} & \begin{minipage}[b]{\linewidth}\raggedright
zk-email \cite{ref9}
\end{minipage} \\
\midrule\noalign{}
\endhead
\bottomrule\noalign{}
\endlastfoot
Privacy & Full ZK & Varies & Attestor sees data & Issuer sees data &
Full ZK & Full ZK \\
Verifiability & On-chain, trustless & Trust issuer & Trust attestor &
Trust issuer & On-chain, trustless & On-chain, trustless \\
Hardware required & None & None & None & None & NFC reader & None \\
Trust anchor & Government CAs & New issuers & KYC provider & Token
issuer & Government & Email providers \\
Existing infrastructure & Billions of certs & Must build new & Requires
provider & Requires issuer & NFC passport & DKIM email \\
Revocation & Trustless CRL in ZK & Issuer registry & Off-chain & Issuer
policy & N/A & N/A \\
Regulatory standing & Legally binding & Unresolved & Provider-dependent
& None & Legally binding & None \\
Time to deploy & 3--6 months & 3--5 years & Months & Months & Months &
Months \\
Front-running defense & Registrant binding & N/A & N/A & N/A & Varies &
Varies \\
\end{longtable}

zk-X509's unique position is the combination of \textbf{no hardware
requirement}, \textbf{government-grade trust} with full certificate
chain verification, \textbf{trustless revocation checking},
\textbf{immediate deployability} (no new issuance infrastructure),
\textbf{legal standing} under existing regulations, and \textbf{full
zero-knowledge privacy}, leveraging an infrastructure base of billions
of existing certificates.

\subsection{Quantitative
Comparison}\label{quantitative-comparison}

zk-X509 measurements were taken on macOS Apple Silicon. Other systems'
values are from their published documentation where available.

\begin{longtable}[]{@{}
  >{\raggedright\arraybackslash}p{(\linewidth - 12\tabcolsep) * \real{0.1096}}
  >{\raggedright\arraybackslash}p{(\linewidth - 12\tabcolsep) * \real{0.1233}}
  >{\raggedright\arraybackslash}p{(\linewidth - 12\tabcolsep) * \real{0.1370}}
  >{\raggedright\arraybackslash}p{(\linewidth - 12\tabcolsep) * \real{0.1644}}
  >{\raggedright\arraybackslash}p{(\linewidth - 12\tabcolsep) * \real{0.1507}}
  >{\raggedright\arraybackslash}p{(\linewidth - 12\tabcolsep) * \real{0.1644}}
  >{\raggedright\arraybackslash}p{(\linewidth - 12\tabcolsep) * \real{0.1507}}@{}}
\toprule\noalign{}
\begin{minipage}[b]{\linewidth}\raggedright
Metric
\end{minipage} & \begin{minipage}[b]{\linewidth}\raggedright
zk-X509
\end{minipage} & \begin{minipage}[b]{\linewidth}\raggedright
zk-email
\end{minipage} & \begin{minipage}[b]{\linewidth}\raggedright
Polygon ID
\end{minipage} & \begin{minipage}[b]{\linewidth}\raggedright
Semaphore
\end{minipage} & \begin{minipage}[b]{\linewidth}\raggedright
zkPassport
\end{minipage} & \begin{minipage}[b]{\linewidth}\raggedright
Worldcoin
\end{minipage} \\
\midrule\noalign{}
\endhead
\bottomrule\noalign{}
\endlastfoot
\textbf{ZK Backend} & SP1 zkVM (RISC-V) & Circom + Groth16 & Circom +
Groth16 & Circom + Groth16 & Noir (Ultra Honk) & Semaphore (zk-SNARKs) \\
\textbf{Constraints / Cycles} & 11.8M cycles (P-256) & 1.26M constraints
(measured) & $\sim$100K--200K constraints (est.) & $\sim$150K
constraints (est.) & N/A & N/A \\
\textbf{Proof Generation} & 102s (CPU, multi-core) & Not reported & Not reported & Not reported &
Not reported & Not reported \\
\textbf{Trusted Setup} & Groth16: circuit-specific ceremony (SP1 Groth16 CRS); PLONK: universal & Required (per-circuit) &
Required & Required & Universal (Ultra Honk) & N/A \\
\textbf{On-Chain Gas} & $\sim$300K (est. Groth16) &
$\sim$300K (est. Groth16) & $\sim$500K (verification, docs) &
$\sim$300K (est.) & $\sim$300K--500K (est. Ultra Honk) &
$\sim$200K (est.) \\
\textbf{Hardware Required} & None & None & None & None & NFC reader &
Orb biometric \\
\textbf{PKI Compatibility} & Any X.509 CA & DKIM (email only) & DID only
& None (custom) & Passport/eID chip & None \\
\textbf{Credential Source} & Government PKI & Email providers & New DID
issuers & None & Passport/eID & Biometric \\
\textbf{Privacy Level} & Full (Merkle CA) & Partial (reveals domain) &
Selective disclosure & Group membership & Full ZK (selective disclosure) & Iris code \\
\textbf{Private Key Isolation} & Yes (key never in circuit) & N/A (no
user private key; email content in circuit) & No & No & Partial (chip
signs externally) & N/A \\
\textbf{Cross-DApp Unlinkability} & Yes (contract-bound nullifier) & No
& Yes & Yes (group-scoped) & Yes (scoped nullifier) & No \\
\textbf{Cross-Chain Replay Defense} & Yes (chain\_id in proof) & No & No
& No & Yes (chainId in BoundData) & No \\
\textbf{Immediate Deployability} & Yes (existing certs) & Yes (existing
email) & No (new DID infra) & Yes & Partial (NFC) & No (Orb) \\
\end{longtable}

\textbf{Key findings:} - \textbf{zk-X509 is the only system} supporting
any X.509 CA worldwide with full CA-membership hiding - \textbf{zk-email} has
comparable on-chain cost but limited to DKIM email signatures (not
government PKI) - \textbf{Polygon ID} requires building entirely new DID
issuance infrastructure - \textbf{zk-X509's private key isolation} is a
an architectural advantage shared only with hardware-dependent systems
like zkPassport (where the passport chip signs externally) --- but
zk-X509 achieves this without any specialized hardware - \textbf{SP1 cycle count (11.8M)} is
higher than Circom constraint counts, but SP1 provides general-purpose
programmability (Rust) vs.~Circom's DSL limitations

\section{Limitations and Future
Work}\label{limitations-and-future-work}

\subsection{Client-Side Proving}\label{client-side-proving}

The current architecture runs the prover as a native application. While the
private key never leaves the local machine (assumption A1), moving proof
generation entirely into the browser via WebAssembly would eliminate
even the inter-process transfer. SP1's WASM support is under active
development and would enable a fully browser-contained proving flow,
strengthening the trust model.

\subsection{On-Chain CRL Oracle}\label{on-chain-crl-oracle}

The current implementation commits a \texttt{crlMerkleRoot} as a public
value, and the smart contract validates it against a stored root
(settable via \texttt{updateCrlMerkleRoot()}). This enables on-chain
enforcement of CRL checking when a trusted operator maintains the CRL
Sorted Merkle Tree. A further improvement would be a fully decentralized
CRL oracle contract maintained by a DAO, automatically fetching and
parsing CRLs from CA distribution points to update the on-chain root
without centralized trust.

\subsection{Multi-Signature
Governance}\label{multi-signature-governance}

The single-owner access control for CA management represents a
centralization point. Replacing it with a multi-signature wallet (e.g.,
Gnosis Safe) and timelock would distribute trust and prevent unilateral
CA whitelist modifications. This is an engineering improvement that does
not affect the core protocol.

\subsection{Cross-Chain Deployment}\label{cross-chain-deployment}

zk-X509 supports multi-chain deployment --- \texttt{IdentityRegistry}
can be deployed on Ethereum, Polygon, Arbitrum, or any EVM-compatible
chain with the same verification key. Users generate a separate proof
per chain, each bound to the target chain's \texttt{chain\_id} and
\texttt{registry\_address}. Two privacy-by-design consequences follow
from the domain separation in Section 3.2:

\begin{enumerate}
\def\labelenumi{\arabic{enumi}.}
\item
  \textbf{Cross-chain replay resistance.} The \texttt{chain\_id} in the
  ownership challenge ensures a proof for Ethereum (chain\_id=1) is
  rejected on Polygon (chain\_id=137).
\item
  \textbf{Cross-chain unlinkability.} The \texttt{registry\_address} in
  the nullifier domain means different deployments produce different
  nullifiers for the same certificate. An observer cannot determine
  whether registrations on two chains belong to the same person --- this
  is a deliberate privacy feature, not a limitation.
\end{enumerate}

If cross-chain identity linkage is desired (e.g., for unified
reputation), the user can voluntarily reveal their nullifiers on both
chains. However, this is an opt-in decision that the protocol does not
enforce, preserving privacy by default.

\subsection{OCSP Support}\label{ocsp-support}

The current implementation supports only Certificate Revocation Lists
(CRLs) for revocation checking. Many modern CAs prefer the Online
Certificate Status Protocol (OCSP, RFC 6960), which provides real-time,
per-certificate revocation status rather than distributing a complete
list of revoked serial numbers. OCSP offers two advantages over CRLs:
(1) lower bandwidth --- a single OCSP response ($\sim$1KB)
versus a full CRL that may contain thousands of entries
($\sim$100KB--1MB), and (2) lower latency --- OCSP responses
are generated on demand rather than on a fixed schedule (typically
24--72 hours for CRLs).

Integrating OCSP into the zkVM circuit presents a distinct challenge.
Unlike CRLs, which are signed by the issuing CA and can be verified
independently, OCSP responses are signed by an OCSP responder that may
use a separate key authorized via the \texttt{id-pkix-ocsp-nocheck}
extension. The zkVM would need to verify: (1) the OCSP response
signature, (2) the responder's certificate chain (potentially a separate
chain from the user's certificate), and (3) the response freshness
(\texttt{thisUpdate}/\texttt{nextUpdate}). This approximately doubles
the signature verification workload, adding $\sim$5.7M cycles
for RSA-2048 or $\sim$3.9M for P-256.

A pragmatic approach combines both mechanisms: CRL verification inside
the zkVM (as currently implemented) for CAs that publish CRLs,
supplemented by on-chain OCSP oracle integration for CAs that primarily
use OCSP. The \texttt{crlMerkleRoot} on-chain validation mechanism
already provides a framework for this --- an OCSP oracle could maintain
a Sorted Merkle Tree of revoked serial numbers derived from OCSP
queries, with the root stored on-chain and validated during proof
verification.

\subsection{Formal Verification}\label{formal-verification}

Formal verification of the Solidity smart contract (e.g., using Certora
or Halmos) and the ZK circuit logic would provide stronger assurance
beyond the game-based security analysis presented here.

\section{Conclusion}\label{conclusion}

zk-X509 demonstrates that legacy PKI infrastructure can be bridged to
blockchain identity systems without compromising user privacy. By
executing full X.509 certificate chain verification---including
multi-level CA signature verification (RSA and ECDSA), temporal
validity, trustless CRL checking, key ownership proof, registrant
binding, configurable multi-wallet policy, and CA-anonymous Merkle
verification---inside a zero-knowledge virtual machine, the system
achieves on-chain verifiability with off-chain privacy. The self-service
re-registration mechanism further ensures that users maintain sovereign
control over their identity lifecycle without centralized admin
dependencies.

The signature-based ownership scheme ensures that the user's private key
never enters the ZK circuit or the prover's general process memory---a
stronger security model in the dimension of private key exposure than
existing ZK identity systems. The
CA Merkle tree design hides which specific CA issued the certificate,
significantly enlarging the anonymity set in multi-national deployments.
The security analysis under the Dolev-Yao model establishes eight
properties with game-based definitions and proofs: unforgeability
(reduced to EUF-CMA security and ZK soundness), unlinkability (reduced
to EUF-CMA security and ZK zero-knowledge), cross-service unlinkability
(reduced to random oracle preimage resistance and ZK zero-knowledge), cross-chain replay resistance
(via chain ID binding and ZK soundness), double-registration
resistance (via deterministic nullifiers and ZK soundness),
front-running immunity (via registrant binding), CA-membership hiding (via
Merkle hiding and ZK zero-knowledge), and non-transferability (reduced
to EUF-CMA security under the local security assumption). The
implementation demonstrates practical feasibility: $\sim$11.8M
SP1 cycles for single-level P-256 verification ($\sim$17.4M for
RSA-2048) and $\sim$300K gas for on-chain registration
(Groth16).

A key differentiator from DID-based approaches is immediacy: while DID
frameworks require years to bootstrap new issuance infrastructure,
zk-X509 leverages government-grade certificates that are already
deployed and legally binding across multiple jurisdictions. The system
supports simultaneous whitelisting of CAs from any nation---Korean NPKI
($\sim$20M users), Estonian eID ($\sim$1.3M
e-residents), German eID, corporate PKI, and beyond---enabling a single
deployment to serve a global user base without cross-border credential
issuance. We believe this ``bridge the existing, don't build from
scratch'' philosophy represents a pragmatic and underexplored direction
in the blockchain identity literature, complementary to rather than
competing with DID-based systems.

\section*{Availability}

Source code will be made publicly available at:
\url{https://github.com/tokamak-network/zk-X509}

\section*{Acknowledgments}

The authors acknowledge the use of large language models, specifically Claude, Gemini, and GitHub Copilot, for assistance in code review, implementation optimization, and editorial refinement of the manuscript.

\bibliographystyle{plain}
\bibliography{references}

\end{document}